\documentclass{article}
\usepackage{amsmath}
\usepackage{bm}
\usepackage{graphicx}
\pdfoutput=1

\begin{document}

\title{Phase reduction approach to synchronization of nonlinear oscillators\footnote{Contemporary Physics {\bf 57}, 188-214 (2016)}}

\author{
Hiroya Nakao\footnote{Tokyo Institute of Technology, Tokyo 152-8552, Japan. E-mail: nakao@mei.titech.ac.jp}
}

\date{}

\maketitle

\begin{abstract}
Systems of dynamical elements exhibiting spontaneous rhythms are found in various fields of science and engineering,
including physics, chemistry, biology, physiology, and mechanical and electrical engineering.
Such dynamical elements are often modeled as nonlinear limit-cycle oscillators.
In this article, we briefly review {\em phase reduction theory}, which is a simple and powerful method for analyzing the synchronization properties of limit-cycle oscillators exhibiting rhythmic dynamics.
Through phase reduction theory, we can systematically simplify the
nonlinear multi-dimensional differential equations describing a limit-cycle oscillator
to a one-dimensional phase equation, which is much easier to analyze.
Classical applications of this theory, i.e., the phase locking of an oscillator to a periodic external forcing and the mutual synchronization of interacting oscillators, are explained.
Further, more recent applications of this theory to the synchronization of non-interacting oscillators induced by common noise and the dynamics of coupled oscillators on complex networks are discussed.
We also comment on some recent advances in phase reduction theory
for noise-driven oscillators and rhythmic spatiotemporal patterns.\\

\end{abstract}

\section{Introduction}

Abundant examples of spontaneous rhythmic oscillations exist in our world, such as chemical reactions, electric circuits, mechanical vibrations, cardiac cells, spiking neurons, flashing fireflies, calling frogs, and the rhythmic walking of animals and robots~\cite{Winfree}-\cite{Shiogai}.
Such rhythmic nonlinear dynamical systems undergoing stable self-sustained periodic oscillations can be considered {\it limit-cycle oscillators}.
In this article, we give an introductory review of a simple and powerful theoretical method for the analysis of limit-cycle oscillators, called {\it phase-reduction theory}, and its applications to the synchronization phenomena of limit-cycle oscillators.

To be more explicit, phase-reduction theory is a mathematical framework that provides us with a reduced description of a nonlinear limit-cycle oscillator.
Under the condition that the perturbations applied to the oscillator are sufficiently weak,
it systematically approximates the dynamics of the oscillator described by a multi-dimensional state variable using only a single {\it phase} variable, where the properties of the oscillator are encapsulated within its {\em natural frequency} and {\em phase response}.
The multi-dimensional dynamical equation describing the oscillator is reduced to a quite simple one-dimensional {\em phase equation}, which can be analyzed in detail and provides important information of various rhythmic phenomena, while allowing the quantitative aspects to be retained.
As a result of these advantages, phase reduction theory is now widely used in science and engineering, being applied not only in the fields of physics and chemistry, but also in biology, mechanical and electrical engineering, robotics, medicine, etc.

Phase reduction theory has a rather long history; by the 1950s, existence of periodic solutions to nonlinear oscillators under perturbation, which is essential for the phase reduction, had already been discussed by Malkin~\cite{Hoppensteadt,Malkin}. In 1960s, Winfree asserted the importance of the notion of phase and formulated the phase model for a population of nonlinear oscillators in his pioneering studies on biological synchronization~\cite{Winfree,Winfree1}.
Since then, many researchers have explored various rhythmic phenomena on the basis of phase reduction theory. Among them, in the 1970--80s, Kuramoto extended Winfree's results and proposed a solvable coupled-oscillator model that exhibits collective synchronization transition, which is now known as the {\em Kuramoto model}~\cite{Kuramoto,Acebron,Strogatz4}, as well as the {\em Kuramoto-Sivashinsky equation}~\cite{Kuramoto,Pikovsky}, which describes nonlinear waves and chaos in oscillatory media.
His theory has played important roles in the analysis of spatiotemporal rhythmic patterns in chemical media, such as the target waves in the Belousov-Zhabotinsky reaction~\cite{Winfree,Kuramoto,Glass}.
The idea of phase reduction was also extensively developed by Ermentrout and Kopell~\cite{Ermentrout1,Kopell1,Kopell2,Ermentrout2}, particularly in relation to the mathematical modeling of neural systems. Ermentrout established a simple and useful method for calculating the {\it phase sensitivity function} (in the terminology of this article), which is called the ``adjoint method'', and which is now used as a standard technique in studies of coupled oscillators~\cite{Ermentrout2,Brown}.

During the period in which research into chaotic dynamical systems was extremely active, the field of the rhythmic phenomena and coupled oscillators was relatively inactive; however, this topic has recently regained considerable attention.
One reason for this renewed interest is that phase reduction theory provides us with a simple and unified framework
to qualitatively and quantitatively treat various real-world rhythmic phenomena
whose detailed information is becoming increasingly accessible as a result of recent progress in experimental techniques.

This article is an introductory review of phase reduction theory.
We provide definitions of the {\em phase} and the {\em phase response}, and derive a {\em phase equation} from the dynamical equation describing a limit-cycle oscillator.
We then discuss a few of classical applications of phase reduction theory; specifically, the phase locking of an oscillator to a periodic external forcing and the mutual synchronization of coupled oscillators.
As more recent examples, we also consider the synchronization of non-interacting oscillators caused by common noise, along with the dynamics of coupled oscillators on complex networks.
In addition, we comment briefly on some recent advances, for example, the application of phase reduction theory to noise-driven oscillators and rhythmic patterns in spatially extended media.

This introductory article aims to provide the main ideas of phase reduction theory and does not cover the vast subject of coupled oscillators.
We do not discuss, for example, detailed analysis of the Kuramoto model of collective synchronization, the various coupled-oscillator models and their intriguing dynamics, and applications to real-world rhythmic phenomena.
Readers who are interested in such topics are advised to consider more detailed discussions provided in the literature.

\section{Limit-cycle oscillators}

\begin{figure}[bt]
	\centering
	\includegraphics[width=\hsize,clip]{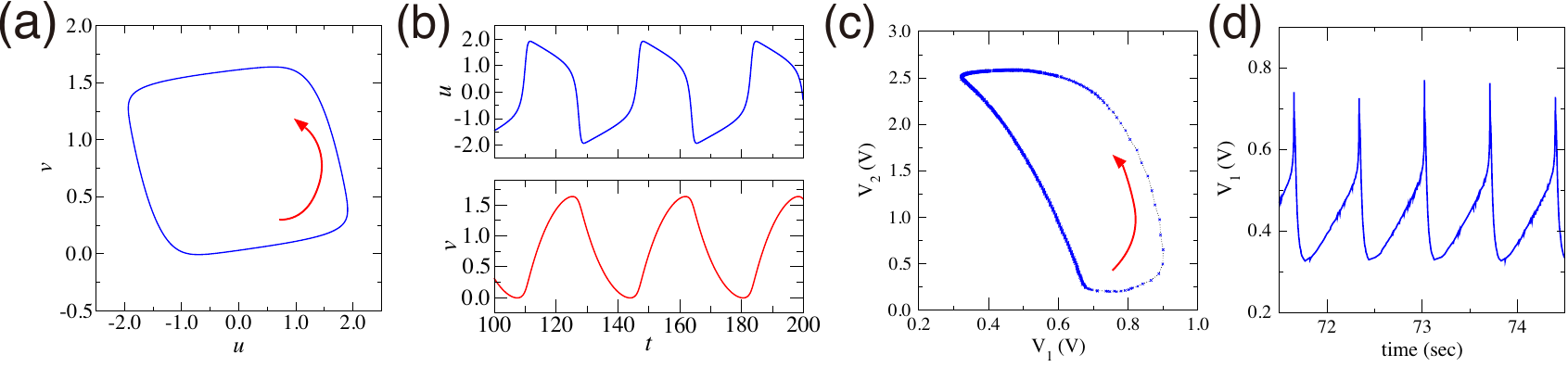}
	\caption{(a,b) Limit-cycle oscillations of the FitzHugh-Nagumo model of a spiking neuron. (a) Dynamics in the state space and (b) time sequences of variables $u$ and $v$. The parameter values are $\epsilon=0.08$, $a=0.7$, $b=0.8$, and $I=0.8$.
	(c,d) Limit-cycle oscillations of an electric circuit that periodically flashes an LED. Two voltage sequences, $V_{1}$ and $V_{2}$, were measured from the circuit via an A/D converter. (c) Dynamics in the $V_{1}$-$V_{2}$ plane and (d) time sequence of $V_{1}$. (Courtesy of K. Arai~\cite{Arai}).}
	\label{fig1}
\end{figure}

In this article, we focus on nonlinear autonomous dynamical systems that exhibit self-sustained steady rhythmic oscillations, calling them simply {\it oscillators}.
More explicitly, we consider a continuous dynamical system possessing a stable limit-cycle orbit in its state space\footnote{In dynamical systems theory, the term ``phase space'' is generally used instead of ``state space''. However, in this article, we use ``state space'' to avoid confusion with the ``phase'' of the oscillations.}.
The periodic dynamics of cardiac cells, the firing of spikes in neurons receiving a constant periodic current~\cite{Winfree,Glass,Hoppensteadt,Tass,Ermentrout1,Schultheiss},
the leg movements of passively walking robots~\cite{McGeer},
and electric circuits powering flashing LEDs (see Fig.~\ref{fig1})~\cite{Arai}
are typical examples of limit-cycle oscillations.
For a more detailed introduction to dynamical systems and rhythmic phenomena, see, e.g.,~\cite{Rosenblum,Janson}, as well as textbooks on dynamical systems (e.g., \cite{Ermentrout1,Strogatz0,Strogatz1,GH,Hale}).

Suppose that the state of the oscillator is represented by $M$ real variables, i.e., it has a $M$-dimensional state space, and denote the oscillator state by a real vector $  {\bf X} = (X_1, \cdots, X_{M})$.
The dynamics of the oscillator is described by an ordinary differential equation (ODE),
\begin{equation}
\frac{d}{dt}{\bf X}(t) = {\bf F}({\bf X}),
\label{ode}
\end{equation}
where $t$ is the time and $ {\bf F}({\bf X}) = (F_{1}({\bf X}), ..., F_{M}({\bf X}))$ is the vector field representing the oscillator dynamics. Here, ${\bf F}({\bf X})$ depends on $t$ only through ${\bf X}$ and does not explicitly depend on $t$, because the oscillator is autonomous.
We assume that this dynamical system has a single stable limit cycle $ {\bf X}_0(t) $ in the state space, which we denote by $\chi$.
(The limit cycle is linearly stable if all of its Floquet exponents have negative real parts, except for a single zero exponent corresponding to the direction along the orbit~\cite{Strogatz1,GH}. See Appendix~\ref{App:Floquet} for a brief explanation.)
The natural period of the limit cycle is denoted by $T$ and the natural frequency is defined as $\omega = 2 \pi / T$, where $ {\bf X}_0(t+T) = {\bf X}_0(t) $ holds. Note that we say ``natural'' here, because the period and frequency of the oscillator may vary when the oscillator is perturbed.
All orbits starting from the basin of attraction~\cite{Strogatz1,GH} of the limit cycle $\chi$ eventually approach $\chi$ and exhibit periodic oscillations.
In general, an analytical expression for ${\bf X}_{0}(t)$ cannot be obtained for nonlinear oscillators,
apart from a few special cases.
See Appendix~\ref{App:SL} for the Stuart-Landau oscillator, for which ${\bf X}_{0}(t)$ can be analytically solved.

For example, the FitzHugh-Nagumo model~\cite{Winfree,Ermentrout1}, which is a simple mathematical model describing the firing dynamics of a neuron, has a two-dimensional state variable $ {\bf X} = (u, v) $ that obeys
\begin{align}
\frac{d}{dt}u(t) = u - \frac{u^{3}}{3} - v + I, \quad \frac{d}{dt}v(t) = \epsilon( u + a - b v ).
\end{align}
Here, $u$ corresponds to the membrane potential of the neuron, while $v$ represents the dynamics of the ion channels on the membrane in a simplified manner. 
This model exhibits stable limit-cycle oscillations within an appropriate range of the parameters $ \epsilon, a, b$, and $I$, where $I$ corresponds to the input current to the neuron.
Figures~\ref{fig1}(a) and (b) show the limit-cycle orbit in the $(u, v)$ state space and the time sequences of $u$ and $v$, respectively.
In this case, the entire state space apart from a single unstable fixed point (inside the limit cycle, not shown) is the basin of attraction of this limit cycle.
The parameter $\epsilon$ is generally small, so the variable $u$ quickly follows $v$ and exhibits slow-fast relaxation oscillations, which consist of slow dynamics along the two branches of the $u$-nullcline on which $du/dt=0$ (given by the curve $v = u - u^{3}/3 + I$) and quick jumps between them~\cite{Ermentrout1}.

Another example, a limit-cycle orbit measured experimentally from an electric circuit for a periodically flashing LED~\cite{Arai}, is shown in Figs.~\ref{fig1}(c) and (d).
The voltage sequences $V_{1}(t)$ and $V_{2}(t)$ were measured at two different locations in the circuit and sampled via an A/D converter.
Figure~\ref{fig1}(c) shows the dynamics of the circuit on the $V_{1}$-$V_{2}$ plane, while 
Fig.~\ref{fig1}(d) shows the periodic time sequence of $V_{1}(t)$.
The spiky relaxation oscillations of $V_{1}$ indicate that the circuit has slow-fast dynamics~\cite{Ermentrout1}.

\section{Phase description of limit-cycle oscillators}

For a stable limit-cycle oscillator, we can introduce a {\it phase} along the limit-cycle orbit and its basin of attraction.
There are various definitions of the oscillator ``phase''. For example, a naive choice would be to define the phase simply as the geometric angle of the state point.
In phase reduction theory, however, the ``asymptotic phase'' is used, which is quite useful in describing the dynamics of limit-cycle oscillators~\cite{Winfree}.
In the following, the asymptotic phase is simply referred to as the ``phase''.
From the definition of this phase, the {\em phase response function} (also known as phase response curve, phase resetting function, or phase resetting curve~\cite{Winfree,Glass,Tass,Ermentrout1}), a fundamental quantity that characterizes the dynamical properties of the oscillator, is naturally defined~\cite{Winfree}-\cite{Hoppensteadt},\cite{Ermentrout1}.
For a weakly perturbed oscillator in particular, the {\em phase sensitivity function}, which gives linear response coefficients of the oscillator phase to applied perturbations, plays an essential role in the derivation of the phase equation.

\subsection{Definition of the phase}

To describe the oscillator state by its phase value, we need to introduce a {\em phase function} $ \Theta({\bf X}) $ that gives a phase value $ \theta = \Theta({\bf X}) $ satisfying $0 \leq \theta < 2\pi$ of the $M$-dimensional oscillator state $ {\bf X} $ in the basin of the limit cycle $\chi$. (The range of the phase is often assumed to be $ [0,1) $ or $ [0,T) $ instead of $[0, 2\pi)$ in the literature. The definition of the frequency also changes accordingly.)
In assigning the phase value to the oscillator state, it is desirable that the resulting phase equation has a simple expression. Such a phase function can be introduced as follows.

\begin{figure}[bt]
	\centering
	\includegraphics[width=0.8\hsize,clip]{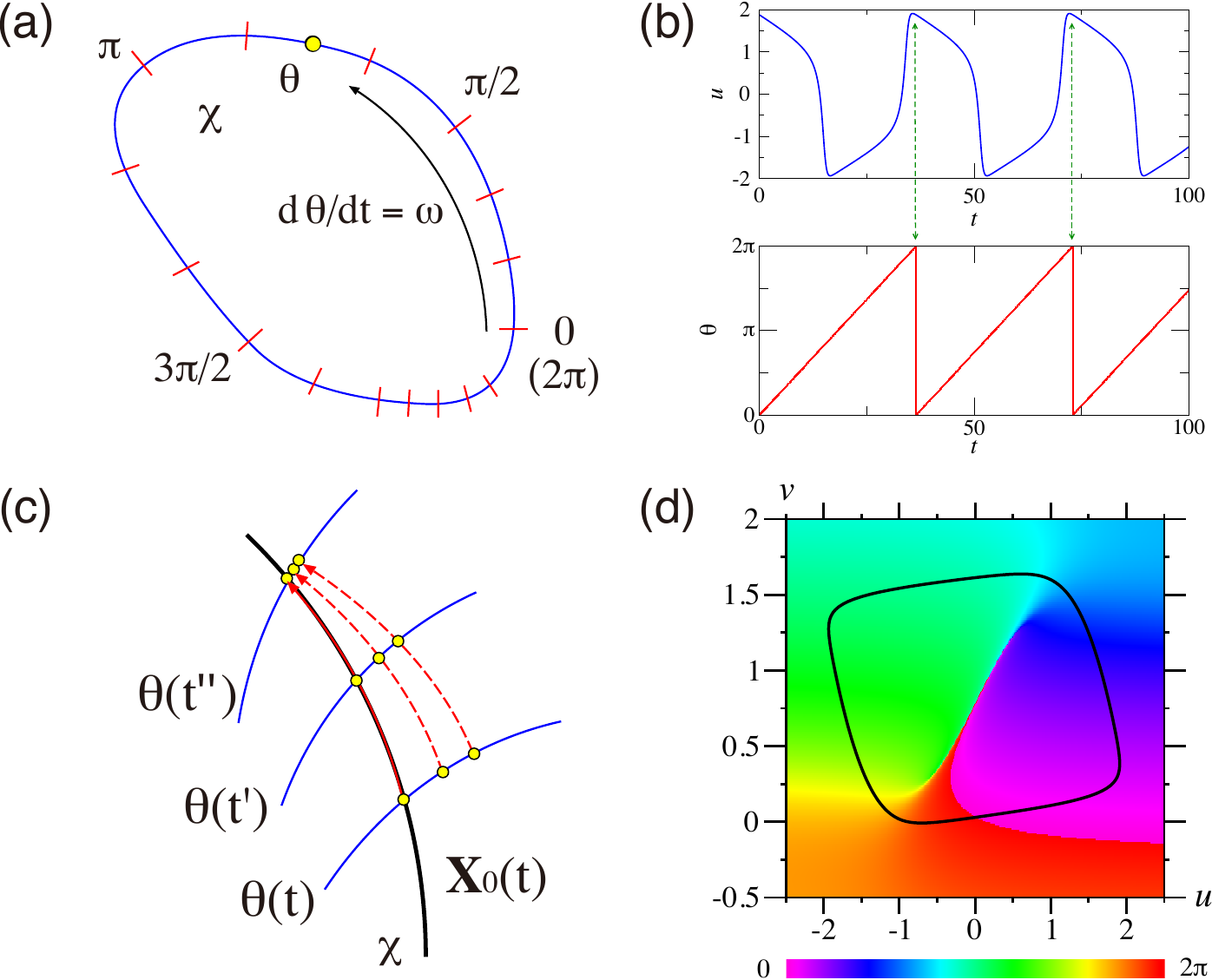}
	\caption{(a)~Definition of the phase of the limit cycle. The phase $\theta$ is identified with the time multiplied by the frequency, i.e., $\theta = \omega t$, yielding a constantly increasing phase variable and non-uniform scales on the limit cycle. (b)~Time series of the variable $u$ and phase $\theta$ of the limit-cycle orbit of the FitzHugh-Nagumo model. (c)~Isochrons (equal-phase sets) of the limit cycle. (d)~Phase function $ \Theta(u,v)$ of the FitzHugh-Nagumo model. In (b) and (d), the parameters of the FitzHugh-Nagumo model are the same as in Fig.~\ref{fig1}.}
	\label{fig2}
\end{figure}

First, we define the phase function $ \Theta({\bf X}) $ {\em on} the limit cycle $\chi$.
As shown schematically in Fig.~\ref{fig2}(a), we choose the origin of the phase $\theta=0$ (identified with $\theta = 2\pi$, as usual) somewhere on $\chi$, and assume that the phase $ \theta(t) = \Theta({\bf X}_{0}(t)) $ of the oscillator state ${\bf X}_{0}(t)$ on $\chi$ increases from $0$ to $2\pi$ as $ {\bf X}_{0}(t) $ goes around $\chi$.
The velocity of the oscillator state ${\bf X}_{0}(t)$ that rotates around $\chi$ is generally {\em not constant} in the state space. However, we can define $\Theta({\bf X})$ so that $\theta(t)$ always increases with {\em a constant frequency} $ \omega = 2\pi / T$.
This is possible by assigning the phase value $\theta(t) = \omega t$ to ${\bf X}_{0}(t)$, in other words, by {\em identifying the phase with the time multiplied by the frequency}.
See Fig.~\ref{fig2}(b) for an example, which shows the time series of the variable $u$ and the phase $\theta$ of the FitzHugh-Nagumo oscillator.
Note that the intervals between the ``scales'' of the phase on the limit cycle are generally not even; they are proportional to the velocity of the oscillator at each state point on $\chi$.
Hereafter, we denote the oscillator state on $\chi$ as a function of $\theta$ (rather than $t$) as ${\bf X}_{0}(\theta)$, where $0 \leq \theta < 2\pi$. Note that $\theta = \Theta({\bf X}_{0}(\theta))$ holds by definition.

When the oscillator is isolated, every oscillator state starting from the basin eventually converges to $\chi$ because the oscillator is autonomous.
However, external perturbations, such as the interaction with other oscillators, may kick the oscillator state out from $\chi$. Therefore, we need to extend the definition of $\Theta({\bf X})$ to the basin of $\chi$.
It is convenient to {\it assign the same $\theta(t)$ to all oscillator states $\{ {\bf X}(t) \}$ that converge to the same ${\bf X}_{0}(\theta(t))$ on $\chi$, which has phase $\theta(t)$, when $t \to \infty$} in the absence of perturbation (see Fig.~\ref{fig2}(c) for a schematic).
This extends the definition of the {\em phase} to the entire basin of $\chi$, as shown in Fig.~\ref{fig2}(d). The set of oscillator states that share the same phase value is called the {\em isochron} (equal-phase set) and plays an important role in the description of limit-cycle oscillations~\cite{Winfree}.
If the oscillator states are on the same isochron at a given point in time, they continue to be on the same isochron for all subsequent time, and converge to the same state on the limit cycle $\chi$.
See Guckenheimer~\cite{Guckenheimer} for mathematical properties of the isochron.

Let us describe the above idea mathematically. By differentiating $\theta(t) = \Theta({\bf X}(t))$ with respect to time, we obtain
\begin{align}
\frac{d}{dt}{\theta}(t) &= \frac{d}{dt}{\Theta}({\bf X}(t))
\cr
&= \mbox{grad}_{{\bf X} = {\bf X}(t)} \Theta({\bf X}) \cdot \frac{d}{dt}{\bf X}(t)  
= \mbox{grad}_{{\bf X} = {\bf X}(t)} \Theta({\bf X}) \cdot {\bf F}({\bf X}(t)),
\end{align}
where the chain rule of differentiation is used and
\begin{align}
\mbox{grad}_{{\bf X} = {\bf X}(t)} \Theta({\bf X})
= \left( \frac{\partial}{\partial X_{1}} \Theta({\bf X}), \cdots, \frac{\partial}{\partial X_{M}}\Theta({\bf X}) \right)_{{{\bf X} = {\bf X}(t)}}
\end{align}
is the gradient of $\Theta({\bf X})$ estimated at ${\bf X} = {\bf X}(t)$.
Therefore, if we define $ \Theta({\bf X}) $ in such a way that
\begin{align}
\mbox{grad}_{\bf X} \Theta({\bf X}) \cdot {\bf F}({\bf X}) = \omega
\label{defphase}
\end{align}
is satisfied at every ${\bf X}$ on $\chi$ and in its basin of attraction, the relation
\begin{align}
\frac{d}{dt} \theta(t) = \omega
\label{phase0}
\end{align}
constantly holds.

Thus, by using the above definition, we can introduce a phase $\theta(t)$ of the oscillator state ${\bf X}(t)$ that increases with a constant $\omega$ in the whole basin of $\chi$.
The multi-dimensional equation~(\ref{ode}) describing the oscillator dynamics is then reduced to a quite simple one-dimensional phase equation~(\ref{phase0}).
Note that we cannot retrieve the full information of ${\bf X}$ from $\theta$ because $\Theta({\bf X})$ is not a one-to-one mapping.
However, if ${\bf X}$ remains {\em near the limit cycle} $\chi$, we can approximate ${\bf X}$ by ${\bf X}_{0}(\theta)$ on $\chi$ with $\theta = \Theta({\bf X})$; this is sufficient to yield the lowest-order phase equation when we consider weakly perturbed oscillators.
This idea is the basis of phase reduction theory.

\subsection{Measurement of the phase}

We saw that the dynamics of the oscillator can be simplified to a one-dimensional phase equation~(\ref{phase0}) by defining the phase function $\Theta({\bf X})$ appropriately.
However, the definition of $\Theta({\bf X})$ in Eq.~(\ref{defphase}) appears to be difficult to use in practice.
Also, it is generally impossible to obtain an analytical expression for $\Theta({\bf X})$ (see Appendix~\ref{App:SL} for a solvable example of the Stuart-Landau oscillator).
In actual numerical simulations or experiments, the oscillator phase can be measured by evolving the oscillator state and calculating the time necessary for the oscillator to reach and pass through the origin of the phase on $\chi$. This is achieved as follows.

Suppose that the initial state of the oscillator is at some point ${\bf X}_{1}$ on $\chi$. To measure the phase of ${\bf X}_{1}$, we evolve the oscillator state from ${\bf X}_{1}$ along the limit cycle.
If time $ \tau\ (0 \leq \tau \leq T) $ is required for the oscillator state to reach and pass through the origin of the phase on $\chi$, the phase $ \theta = \Theta({\bf X}_{1}) $ of ${\bf X}_{1}$ is given by $ \theta = 2 \pi - \omega \tau = 2 \pi ( 1 - \tau / T ) $, because $ \theta + \omega \tau = 2 \pi $.
Next, to measure the phase of the initial state ${\bf X}_{2}$ in the basin of $\chi$ (but not exactly on $\chi$), we evolve the oscillator state from ${\bf X}_{2}$ for time $n T$, where $n$ is some positive integer, until it converges sufficiently close to $\chi$.
If the state converges to an oscillator state with phase $\theta$ on $\chi$, the phase of ${\bf X}_{2}$ is also $\theta$ (the phase is considered in modulo $2\pi$ here).

In Fig.~\ref{fig2}(d), the phase function $ \Theta(u,v) $ of the FitzHugh-Nagumo model, which was calculated numerically using the above algorithm, is shown on the $(u,v)$ state space.
Reflecting the slow-fast dynamics of this model, the phase function largely varies along the left and right branches of the limit-cycle orbit where the oscillator state slowly follows the $u$-nullcline, while it shows little variation along the upper and lower branches where the oscillator state quickly jumps between the left and right branches. The $u$-nullcline clearly separates the phase portrait, because oscillator states to the left of this curve are quickly attracted to the left branch, and those to the right are attracted to the right branch.
See Fig.~\ref{figA1} in Appendix~\ref{App:SL} for comparison, which shows the phase function of a smooth Stuart-Landau oscillator that varies smoothly along the limit-cycle orbit.

In experiments, only some of the state variables of the oscillator may be observed. However, provided we have sufficiently stable limit-cycle oscillations and can detect that the oscillator state passes through the origin of the phase on $\chi$ in some way, we can, in principle, measure the oscillator phase even from a single time sequence of the oscillator.
In extracting the phase information from experimental signals, linear interpolation of threshold-crossing events, wavelet transform, or Hilbert transform of the measured signal is typically used~\cite{Pikovsky,Shiogai}. The resulting phase (called ``protophase'' in Refs.~\cite{Kralemann,Schwabedal,Kralemann2}) is generally different from the asymptotic phase used in phase reduction theory, but it can be further converted to the true asymptotic phase that increases with a constant frequency by nonlinear transformation of the variables~\cite{Kralemann,Schwabedal,Kralemann2}.

\begin{figure}[bt]
	\centering
	\includegraphics[width=\hsize,clip]{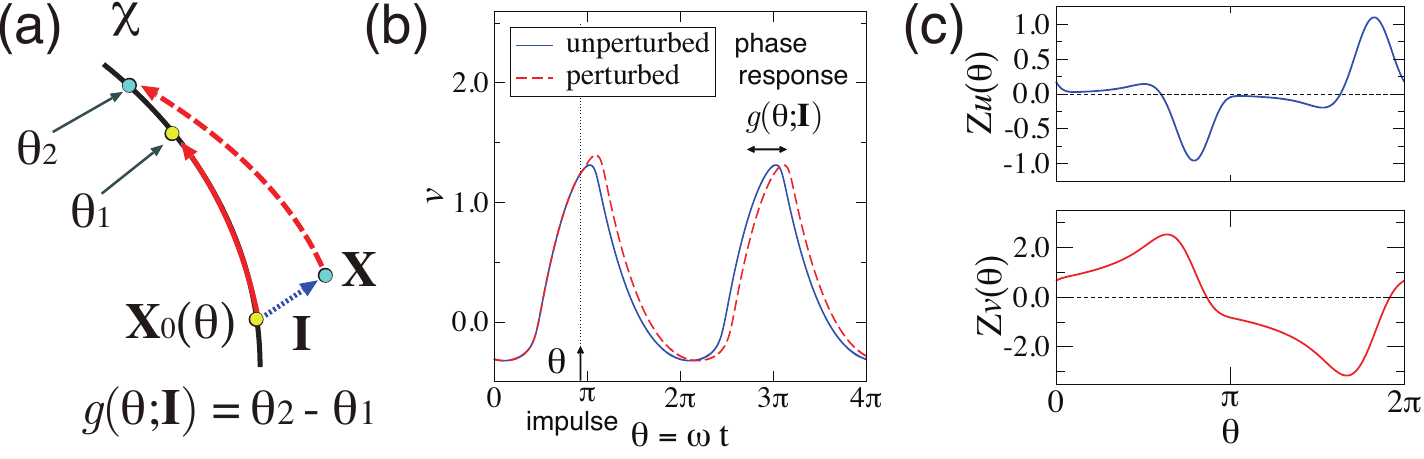}
	\caption{(a),(b)~Phase response of the oscillator to an impulsive stimulus (state space and time sequence). (c)~Phase sensitivity function ${\bf Z}(\theta) = (Z_{u}(\theta), Z_{v})(\theta)$ of the FitzHugh-Nagumo model. The parameters are the same as in Fig.~\ref{fig1}.}
	\label{fig3}
\end{figure}

\subsection{Phase response to external perturbations}

We have defined the phase and derived the phase equation for an isolated oscillator.
However, we must also incorporate perturbations applied to the oscillator into the phase equation, because the oscillators are usually driven by external forcing or are mutually interacting (otherwise no interesting phenomena will occur!).
In phase reduction theory, the response function of the oscillator phase to external
perturbations plays an important role.

Let us introduce the phase response of the oscillator to an impulsive stimulus. Suppose that the oscillator is perturbed by an impulse, represented by a perturbation vector $ {\bf I} = (I_{1}, ..., I_{M})$, when its state is at ${\bf X}_0(\theta) $ with phase $\theta$ on the limit cycle $\chi$ (see Fig.~\ref{fig3}(a) for a schematic).
We assume that the oscillator is kicked to a new state, $ {\bf X} = {\bf X}_0(\theta) + {\bf I} $, within the basin of $\chi$ (if ${\bf I}$ is too strong and the state is expelled from the basin of $\chi$, the state does not return to $\chi$ and phase reduction theory is no longer applicable).
As previously, the phase of the kicked state ${\bf X}$ is given by $ \Theta({\bf X}) $, and this state gradually returns to $\chi$.
By this kick, a phase shift from that of the unperturbed orbit is produced. Using the phase function, this shift can be represented as
\begin{align}
	g(\theta ; {\bf I}) = \Theta({\bf X}_0(\theta) + {\bf I}) - \theta,
\end{align}
because the new phase of the oscillator just after the kick is given by $\Theta({\bf X}_0(\theta) + {\bf I})$.
We call this $ g(\theta ; {\bf I}) $ the {\em phase response function} of the oscillator, which is $2\pi$-periodic in $\theta$ and generally depends nonlinearly on ${\bf I}$.
This function characterizes the response property of the oscillator phase to an external impulsive stimulus given at phase $\theta$ on the limit cycle $\chi$, and is essentially important in phase reduction theory.

In this article, we focus on weakly perturbed oscillators (see Appendix~\ref{App:Impulse} for the treatment of oscillators driven by finite-intensity impulses).
Therefore, we can work in the linear-response framework as follows. When $|{\bf I}|$, the magnitude of ${\bf I}$, is sufficiently small, the phase function can be expanded in a Taylor series as 
\begin{align}
\Theta({\bf X}_{0}(\theta) + {\bf I}) = \Theta({\bf X}_{0}(\theta)) + \left. \mbox{grad}_{\bf X} \Theta(\bf X) \right|_{{\bf X} = {\bf X}_0(\theta)} \cdot {\bf I} + O(|{\bf I}|^{2}),
\end{align}
where $\Theta({\bf X}_{0}(\theta)) = \theta$ by definition.
Therefore, the phase response function can be linearly approximated as
\begin{align}
g(\theta ; {\bf I}) \simeq  \left. \mbox{grad}_{\bf X} \Theta(\bf X) \right|_{{\bf X} = {\bf X}_0(\theta)} \cdot {\bf I}
= {\bf Z}(\theta) \cdot {\bf I},
\label{phaseresponsegz}
\end{align}
where we express the linear response coefficient of the phase to small ${\bf I}$ as
\begin{align}
{\bf Z}(\theta) = \left. \mbox{grad}_{\bf X} \Theta(\bf X) \right|_{{\bf X} = {\bf X}_0(\theta)}.
\end{align}
We call this ${\bf Z}(\theta)$ the {\em phase sensitivity function}, which gives a $M$-dimensional gradient vector of $ \Theta({\bf X}) $ estimated at ${\bf X} = {\bf X}_{0}(\theta)$ with phase $\theta$ on the limit cycle $\chi$.
Equation~(\ref{phaseresponsegz}) indicates that the projection of the perturbation ${\bf I}$ onto the direction along $\chi$, which is given by the inner product of ${\bf I}$ and ${\bf Z}(\theta)$, eventually remains and affects the phase dynamics of the oscillator.

In Fig.~\ref{fig3}(c), $ {\bf Z}(\theta) = ( Z_u(\theta),\ Z_v(\theta) ) $ of the FitzHugh-Nagumo model is shown, which represents the sensitivity of the oscillator phase $\theta$ to weak perturbations applied to $u$ and $v$, as a function of the oscillator phase $\theta$ at which the perturbation is given.
We can see how the oscillator phase is advanced or delayed depending on the timing of the perturbation.

In phase reduction theory, the dynamical property of the oscillator is encapsulated into $ g(\theta ; {\bf I}) $ and, for weakly perturbed oscillators in particular, $ {\bf Z}(\theta) $ plays an essential role in the derivation of the phase equation.
We use the term phase sensitivity function for ${\bf Z}(\theta)$ in this article, because the term ``sensitivity function'' was used for ${\bf Z}(\theta)$ in Winfree's original paper on biological rhythms in 1967~\cite{Winfree1}.
This function is also called the ``phase response function'' or ``infinitesimal phase resetting curve'' in the literature.
Though the functions $g(\theta ; {\bf I})$ and ${\bf Z}(\theta)$ are called in various ways, in any case, $g(\theta ; {\bf I})$ representing the {\em phase shift} due to perturbations should not be confused with ${\bf Z}(\theta)$ representing {\em linear response coefficients} of the phase to perturbations.

To measure the phase response function $ g(\theta ; {\bf I}) $ directly, we stimulate the oscillator by an impulse ${\bf I}$ at varying values of phase $\theta$ on $\chi$, and measure the resulting phase difference from the unperturbed case sufficiently after the transient (see Fig.~\ref{fig3}(b)). 
The phase sensitivity function $ {\bf Z}(\theta) = (Z_{1}(\theta), ..., Z_{M}(\theta))$ can thus be obtained, in principle, by applying a very weak impulsive stimulus of intensity $I$ to each component $X$ of the state variable and measuring $g(\theta ; I {\bf e}_{j})$ as
\begin{align}
Z_{j}(\theta) = \lim_{I \to 0} \frac{ g(\theta ; I {\bf e}_{j}) }{ I },
\end{align}
where ${\bf e}_{j}$ is a unit vector along the $j$th component of the oscillator state ($j=1, ..., M$).
This is sometimes called the {\em direct method} for measurement of the phase response properties.

In experiments, it is not always possible to apply an appropriate stimulus to every component of the state variable. Then, the corresponding component of ${\bf Z}(\theta)$ cannot be determined using the direct method.
Also, it is necessary to apply a stimulus with an appropriate intensity so that the response of the oscillator is not buried in experimental noise.
However, because the linear approximation of $g(\theta ; {\bf I})$ is valid only for sufficiently small ${\bf I}$, the resulting ${\bf Z}(\theta)$ can be inaccurate.
Various methods for measuring the phase response and phase sensitivity functions have been devised and applied to experimental data of various kinds of nonlinear oscillators, such as electric circuits, firing neurons, and circadian rhythms~\cite{Schultheiss},\cite{Arai},\cite{Galan}-\cite{Fukuda}.
When the mathematical model of the oscillator is explicitly given, the phase sensitivity function ${\bf Z}(\theta)$ can be more easily and accurately calculated using the ``adjoint method'', as we explain next.

\subsection{The adjoint method and Malkin's theorem}

It is known that the phase sensitivity function ${\bf Z}(\theta)$ can be obtained as a $2\pi$-periodic solution to the following ``adjoint equation''~\cite{Hoppensteadt,Ermentrout1,Ermentrout2}:
\begin{align}
\omega \frac{d}{d\theta} {\bf Z}(\theta) = -{\rm J}(\theta)^{\dag} {\bf Z}(\theta)
\label{adjoint}
\end{align}
with the normalization condition
\begin{align}
	{\bf Z}(\theta) \cdot \frac{d{\bf X}_{0}(\theta)}{d\theta} = 1
	\label{normalization}
\end{align}
for $0 \leq \theta < 2\pi$. Here, ${\rm J}(\theta)$ is the Jacobi matrix of ${\bf F}({\bf X})$ at ${\bf X} = {\bf X}_{0}(\theta)$, i.e., its $(i,j)$-component is given by
\begin{align}
{\rm J}_{ij}(\theta) = \left. \frac{\partial F_{i}({\bf X})}{\partial X_{j}} \right|_{{\bf X} = {\bf X}_{0}(\theta)}
\end{align}
for $1 \leq i,j \leq M$, and $^{\dag}$ indicates the matrix transpose.
Equation~(\ref{adjoint}) is adjoint to the linearized variational equation\footnote{Recall that the variational equation is given by $d{\bf u}(t)/dt = {\rm J}({\bf X}_{0}(t)){\bf u}(t)$ where ${\rm J}({\bf X}_{0}(t))$ is the Jacobi matrix of ${\bf F}({\bf X})$ at ${\bf X} = {\bf X}_{0}(t)$ (Appendix~\ref{App:Floquet}). When $\bf u$ is expressed as a function of $\theta = \omega t$, this gives Eq.~(\ref{variational}).}
\begin{align}
	\omega \frac{d}{d\theta} {\bf u}(\theta) = {\rm J}(\theta) {\bf u}(\theta)
	\label{variational}
\end{align}
describing a small variation ${\bf u}(\theta)$ of ${\bf X}$ from ${\bf X}_{0}(\theta)$, expressed here as a function of $\theta$ rather than $t$, with respect to the inner product defined as
\begin{align}
[{\bf Z}(\theta), {\bf u}(\theta)] = \frac{1}{2\pi} \int_{0}^{2\pi} {\bf Z}(\theta) \cdot {\bf u}(\theta) d\theta.
\end{align}
See Appendix~\ref{App:adjoint} for a derivation of the adjoint equation.

As explained in Ref.~\cite{Hoppensteadt}, the adjoint equation~(\ref{adjoint}) is an essential part of Malkin's theorem, which guarantees the phase description itself~\cite{Hoppensteadt,Ermentrout1,Malkin}.
The phase equation can be seen as a ``solvability condition'' of the Fredholm theorem for the variational equation of periodically perturbed oscillators~\cite{Hoppensteadt,Ermentrout1,Hale}.
When the phase equation is satisfied with ${\bf Z}(\theta)$ given by Eq.~(\ref{adjoint}), the variational equation has a non-divergent periodic solution and the oscillator state can be represented by the phase.
As shown by Ermentrout~\cite{Ermentrout1,Ermentrout2}, the adjoint equation~(\ref{adjoint}) also gives a simple and useful numerical algorithm for calculating ${\bf Z}(\theta)$.
That is, by integrating Eq.~(\ref{adjoint}) backward in time, only the neutrally stable periodic component corresponding to ${\bf Z}(\theta)$ can be obtained.
The phase sensitivity function ${\bf Z}(\theta) = (Z_{u}(\theta), Z_{v}(\theta))$ in Fig.~\ref{fig3}(c) was calculated using the adjoint method. See Appendix~\ref{App:Numerics} for the numerical algorithm.

\subsection{Phase reduction}

Using the phase sensitivity function ${\bf Z}(\theta)$, we can reduce the multi-dimensional dynamical equation for a limit-cycle oscillator driven by weak perturbations to an approximate one-dimensional phase equation~\cite{Winfree,Kuramoto}\footnote{The dynamics of the oscillator receiving impulsive perturbations can also be reduced to a phase equation using the phase response function $g(\theta ; {\bf I})$. In this case, a one-dimensional map for the phase can be derived. The impulses need not be weak, provided the intervals between them are sufficiently large for the oscillator state to return to the limit cycle. See Appendix~\ref{App:Impulse} for a brief explanation.}.
Suppose that the dynamics of the oscillator is described by
\begin{align}
\frac{d}{dt}{\bf X}(t) = {\bf F}({\bf X}) + \varepsilon {\bf p}({\bf X}, t).
\label{oscperturb}
\end{align}
Here, ${\bf F}({\bf X})$ represents the unperturbed oscillator dynamics, $ {\bf p}({\bf X}, t) $ is a $M$-dimensional vector representing the weak perturbation that depends on the oscillator state ${\bf X}$ and the time $t$, and $\varepsilon$ is a {\em small parameter} representing the intensity of the perturbation.
We assume that this $\varepsilon$ is sufficiently small so that the oscillation of the system persists and the perturbed orbit does not deviate significantly from the unperturbed limit cycle $\chi$.

Let us now derive the approximate phase equation for the weakly perturbed oscillator. First, using the phase function $\Theta({\bf X})$, we can rewrite the phase dynamics as
\begin{align}
\frac{d}{dt}{\theta}(t) 
&= \frac{d}{dt}{\Theta}({\bf X}(t)) 
= \left. \mbox{grad}_{\bf X} \Theta({\bf X})\right|_{{\bf X}={\bf X}(t)} \cdot \frac{d}{dt}{\bf X}(t) \cr
&
= \left. \mbox{grad}_{\bf X} \Theta({\bf X})\right|_{{\bf X}={\bf X}(t)} \cdot \left\{ {\bf F}({\bf X}(t)) + \varepsilon {\bf p}({\bf X}(t), t)
\right\}
\cr
&= \omega + \varepsilon \left. \mbox{grad}_{\bf X} \Theta({\bf X})\right|_{{\bf X}={\bf X}(t)} \cdot {\bf p}({\bf X}(t), t),
\label{phasenotclosed}
\end{align}
where we used Eq.~(\ref{defphase}).
Though no approximation is made at this point, this equation still depends explicitly on ${\bf X}(t)$ and is not closed in $\theta$ because of the last perturbation term.

To obtain a phase equation that is closed in $\theta$, we use the fact that the perturbation is sufficiently weak and, therefore, {\em the oscillator state does not deviate significantly from the unperturbed limit cycle $\chi$}.
Thus, at the lowest order, we may approximate ${\bf X}(t)$ in the above equation by a nearby state ${\bf X}_{0}(\theta(t))$ on $\chi$ that has the same phase value $\theta(t) = \Theta({\bf X}(t))$ as ${\bf X}(t)$.
Then, the term 
$\left. \mbox{grad}_{\bf X} \Theta({\bf X})\right|_{{\bf X}={\bf X}(t)} \cdot {\bf p}({\bf X}(t), t) $
in Eq.~(\ref{phasenotclosed}) can be approximated by
$ \left. \mbox{grad}_{\bf X} \Theta({\bf X}) \right|_{{\bf X} = {\bf X}_0(\theta(t))} \cdot {\bf p}({\bf X}_0(\theta(t)), t) = {\bf Z}(\theta(t)) \cdot {\bf p}({\bf X}_0(\theta(t)), t)$.
Therefore, Eq.~(\ref{phasenotclosed}) is approximated to a {\em phase equation closed in $\theta$},
\begin{align}
\frac{d}{dt}{\theta}(t) 
= \omega + \varepsilon
{\bf Z}(\theta) \cdot {\bf p}(\theta, t),
\label{phseq}
\end{align}
where $ {\bf p}(\theta, t) = {\bf p}({\bf X}_0(\theta), t) $ represents the perturbation given at the oscillator state ${\bf X} = {\bf X}_{0}(\theta)$ on the limit cycle. 
The error yielded by this approximation is generally $O(\varepsilon^{2})$,
because the replacement of ${\bf X}(t)$ by ${\bf X}_{0}(\theta(t))$ in
$\mbox{grad}_{\bf X} \Theta({\bf X})$ yields an error of $O(\varepsilon)$.

Thus, up to the first order in $\varepsilon$, we can obtain an approximate equation for the phase $\theta$ describing a weakly perturbed oscillator in a closed form. The original {\em multi-dimensional} dynamical equation~(\ref{oscperturb}) describing a perturbed oscillator has been systematically reduced to a simple {\em one-dimensional} phase equation~(\ref{phseq}), which is much easier to analyze.
In the following, we use this reduced phase equation to analyze several types of synchronization phenomena.

\section{Synchronization of oscillators due to external forcing}

In this section, we analyze two kinds of synchronization dynamics of limit-cycle oscillators driven by weak external forcing, using phase reduction theory. The first case is a classical example of the phase locking of oscillators to periodic external forcing. The second case is a relatively new example of noise-induced phase synchronization occurring between non-interacting oscillators, which is caused by common noisy forcing. Both situations can be analyzed in a general manner using phase reduction theory.

\subsection{Periodic forcing}

As the first example, we consider the phase locking of limit-cycle oscillators to periodic external forcing. This is the most basic synchronization phenomenon and has therefore been studied extensively.
For example, our circadian rhythm is phase-locked to the $24$h day-night cycle of the sun~\cite{Winfree}. The injection locking of electric circuits and lasers using periodic signals has been used in engineering~\cite{York,Harada,Zlotnik}.
Here, we consider the phase locking of nonlinear oscillators driven by weak periodic forcing.

Suppose that an oscillator is weakly driven by a periodic external forcing ${\bf f}(t)$ of period $T_{ext}$ and frequency $\Omega = 2 \pi / T_{ext}$, which satisfies ${\bf f}(t+T_{ext}) = {\bf f}(t)$.
The oscillator obeys
\begin{align}
\frac{d}{dt}{\bf X}(t) = {\bf F}({\bf X}) + \varepsilon {\bf f}(t),
\end{align}
where $\varepsilon$ is a small parameter representing the forcing intensity.
We consider the external forcing $\varepsilon {\bf f}(t)$ as a weak perturbation and plug ${\bf p}(\theta, t) = {\bf f}(t)$ into Eq.~(\ref{phseq}). Then, the above equation for ${\bf X}(t)$ can be approximated to a phase equation of the form
\begin{align}
\frac{d}{dt}{\theta}(t) = \omega + \varepsilon {\bf Z}(\theta) \cdot {\bf f}(t).
\label{phaseperiodic}
\end{align}

\begin{figure}[bt]
        \centering
       	\includegraphics[width=\hsize,clip]{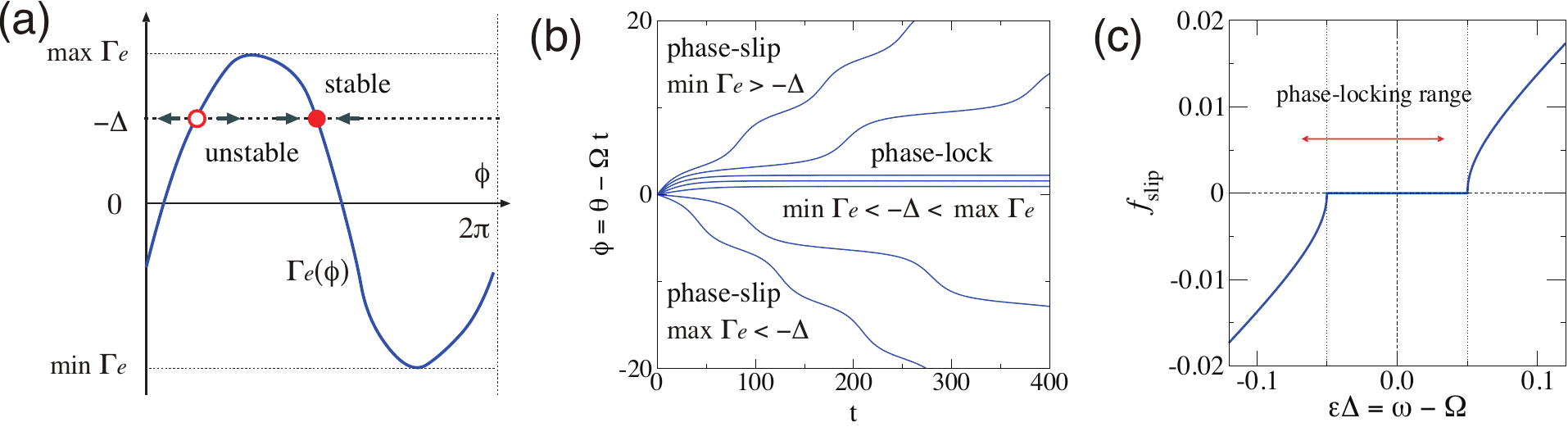}
        \caption{Dynamics of the relative phase of an oscillator driven by periodic external forcing. (a) Dynamics of the relative phase $\phi(t) = \theta(t) - \Omega t$ and the function $\Gamma_{e}(\phi)$.
        (b) Time evolution of the relative phase $\phi(t)$. When the frequency difference is sufficiently small and Eq.~(\ref{phaselockcond}) is satisfied, phase locking takes place and $\phi(t)$ converges to a constant.
        (c) Frequency of the phase slips $f_{{slip}}$ vs. the frequency difference $\varepsilon \Delta = \omega - \Omega$. When $|\varepsilon \Delta|$ is sufficiently small, phase locking takes place and phase slips do not occur, i.e., $f_{{slip}}=0$. Outside this range, the relative phase continues to increase or decrease and $f_{{slip}} \neq 0$.
		In (c), it is assumed that the phase sensitivity function is $Z(\phi) = \sin \phi$, the periodic forcing is $f(t) = \sin \Omega t$, and the intensity of the perturbation is $\varepsilon = 0.1$, which yields $\Gamma_{e}(\phi) = ( \cos \phi ) / 2$, $\mbox{min } \Gamma_{e} = -1/2$, $\mbox{max } \Gamma_{e} = 1/2$, and the phase-locking range $-0.05 < \varepsilon \Delta < 0.05$.}
        \label{fig-injection}
\end{figure}

Although reduced to a phase-only equation, Eq.~(\ref{phaseperiodic}) is still {\em non-autonomous}, that is, the right-hand side depends explicitly on $t$.
By applying the {\em averaging method}~\cite{Kuramoto,Hoppensteadt,Sanders}, we can further simplify this equation to an {\it autonomous} phase equation for $\theta$, which can more easily be analyzed~\cite{Kuramoto}.
To this end, we assume that both frequencies, i.e., $\Omega$ and $\omega$, are sufficiently close.
More explicitly, we assume that the frequency difference $\omega - \Omega$ is $O(\varepsilon)$, and denote it as 
\begin{align}
\omega - \Omega = \varepsilon \Delta,
\end{align}
where $\Delta$ is $O(1)$.
Namely, we consider the case that the frequency difference and the periodic forcing are of the same order.
We introduce a relative phase,
\begin{align}
\phi(t) = \theta(t) - \Omega t,
\end{align}
which is the difference between $\theta(t)$ and the phase of the external forcing $\Omega t$.
Then, from Eq.~(\ref{phaseperiodic}), $\phi(t)$ obeys
\begin{align}
\frac{d}{dt}{\phi}(t)
 = \omega - \Omega + \varepsilon {\bf Z}(\phi(t) + \Omega t) \cdot {\bf f}(t) 
 = \varepsilon \{ \Delta + {\bf Z}(\phi(t) + \Omega t) \cdot {\bf f}(t) \}.
\end{align}

Now, because the right-hand side of the above equation is $O(\varepsilon)$, $\phi(t)$ is a slowly varying variable, while $\Omega t$ varies rapidly.
Thus, we may approximate the right-hand side of the above equation by integrating it over one period of the fast external forcing, assuming that $\phi(t)$ does not vary within $T_{ext}$.
By this averaging approximation\footnote{More precisely, $\phi$ (the phase before averaging) and $\tilde{\phi}$ (the phase after averaging) are slightly different. They are related by a near-identity transform of the form $\phi = \tilde{\phi} + \varepsilon h(\tilde{\phi},t)$~\cite{Hoppensteadt,Sanders}. Although we treat $\phi$ and $\tilde{\phi}$ as being identical in this article, their dynamics are slightly different. For example, even if $\tilde{\phi}$ becomes constant and the oscillator is ``phase-locked'' to the periodic forcing, $\phi$ can exhibit tiny oscillations around the locked phase.}, we obtain
\begin{equation}
\frac{d}{dt}{\phi}(t) = \varepsilon \{ \Delta + \Gamma_{e}(\phi) \}.
\label{phsperiodsingle}
\end{equation}
Here, $ \Gamma_{e}(\phi) $ is a $2\pi$-periodic function representing the effect of the periodic external forcing on the oscillator phase, and is given by
\begin{align}
\Gamma_{e}(\phi)
=
\frac{1}{T_{ext}} \int_{t}^{t+T_{ext}}
 {\bf Z}(\phi + \Omega t') \cdot {\bf f}(t') dt'
=
\frac{1}{2\pi} \int_{0}^{2\pi}
{\bf Z}(\phi + \psi) \cdot {\bf f}\left(\frac{\psi}{\Omega}\right) d\psi
\end{align}
using the $2\pi$-periodicity of ${\bf Z}(\theta)$ and the $T_{ext}$-periodicity of ${\bf f}(t)$.
The error caused by this approximation is also $O(\varepsilon^{2})$~\cite{Hoppensteadt}.

Because Eq.~(\ref{phsperiodsingle}) is a one-dimensional {\em autonomous} equation, its dynamics can be easily analyzed by drawing a graph of $\Gamma_{e}(\phi)$, as shown schematically in Fig.~\ref{fig-injection}(a).
As we can see, if the condition
\begin{align}
\mbox{min}\ \Gamma_{e}(\phi) < - \Delta < \mbox{max}\ \Gamma_{e}(\phi)
\label{phaselockcond}
\end{align}
is satisfied, Eq.~(\ref{phsperiodsingle}) has at least two fixed points at which $ d{\phi}(t)/dt = 0$ holds, one of which is stable.
Thus, the relative phase $\phi(t) = \theta(t) - \Omega t$ converges to the stable fixed point.
This means that the phase difference between the oscillator and the external forcing becomes constant, that is, the oscillator exhibits {\it phase locking} to the periodic external forcing.
In contrast, if $ |\Delta| $ is too large, $ d{\phi}(t)/dt > 0 $ or $ d{\phi}(t)/dt < 0 $ always hold and $\phi(t)$ continues to increase or decrease. The oscillator cannot phase lock to the external forcing in this case; it exhibits periodic phase slips and the relative phase increases indefinitely.
See Fig.~\ref{fig-injection}(b) for the dynamics of the relative phase.

The mean frequency of the oscillator is equal to $\Omega$ when the oscillator is phase-locked to the external forcing, while it deviates from $\Omega$ when the oscillator exhibits phase slips.
Figure~\ref{fig-injection}(c) shows the frequency $f_{{slip}} = 1 / T_{{slip}}$ of the phase slips, where $T_{{slip}}$ is the time interval between slips, as a function of the frequency difference $\varepsilon \Delta = \omega - \Omega$.
When phase locking takes place, the mean oscillator frequency equals $\Omega$ and $f_{{slip}} = 0$. Outside this range, phase slips occur and $f_{{slip}}$ takes either positive or negative values. In this case, $f_{{slip}}$ can be calculated as
\begin{align}
f_{{slip}} = \frac{1}{T_{{slip}}},
\quad
T_{{slip}} = \int_{0}^{2\pi} \frac{d\phi}{\varepsilon \{ \Delta + \Gamma_{e}(\phi) \}},
\end{align}
from Eq.~(\ref{phsperiodsingle}).
Thus, the graph of $f_{{slip}}$ has a characteristic plateau in the phase-locking range as shown in Fig.~\ref{fig-injection}(c).

The above argument can be generalized to the case where the ratio of $\omega$ to $\Omega$ is close to a rational value, i.e., $\omega / \Omega \approx n / m$ with integer $n$ and $m$, and it can be shown that the oscillator generally exhibits $m$:$n$ phase locking to the external forcing, i.e., the oscillator rotates exactly $m$ times while the external forcing oscillates $n$ times.
In principle, the phase locking occurs for every pair of $m$ and $n$, i.e., when the natural frequency $\omega$ of the oscillator is close to rational multiples of $\Omega$. If the mean frequency of the driven oscillator is plotted as a function of $\Omega$, this leads to the well-known Devil's staircase graph with many plateaus of various sizes~\cite{Glass,Ermentrout1}.

\subsection{Common noisy forcing}

As the second example, we consider synchronization of non-interacting oscillators induced by common weak noise.
It is known that dynamical systems tend to exhibit synchronized or coherent behavior when they are driven by common noise, even if no direct interaction exists between them.
Similarly, if a single dynamical system is repetitively driven by the same sequence of noise, it tends to reproduce the same response even if the initial conditions are different among trials.
Some examples include the synchronization of chaotic oscillators driven by common random signals~\cite{Sanchez,Toral,Zhou}, improvements in the reproducibility of the dynamics of lasers~\cite{Uchida} and the firing timing of neurons~\cite{Mainen,Galan2} receiving random signals, and synchronized variations in wild animal populations located in different, distant areas caused by common environmental fluctuations~\cite{Grenfell}.

\begin{figure}[bt]
        \centering
	   	\includegraphics[width=\hsize,clip]{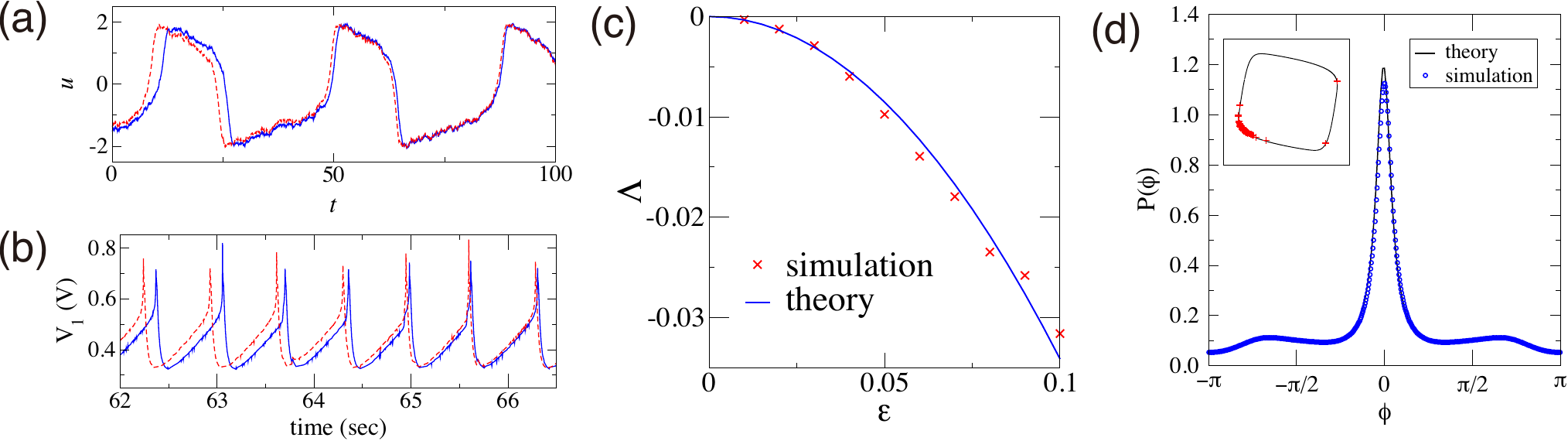}
        \caption{(a) Synchronization of two non-interacting FitzHugh-Nagumo oscillators driven by common white noise. (b) Reproducibility of a flashing LED driven by the same sequence of random impulses. (c) Lyapunov exponent of the FitzHugh-Nagumo oscillator driven by Gaussian-white noise. Theoretical values obtained from $Z(\theta)$ are compared with numerical values obtained by direct numerical simulations of the noise-driven FitzHugh-Nagumo oscillator. (d) Stationary probability density function of the phase differences of $200$ FitzHugh-Nagumo oscillators driven by common and independent noise terms. The inset shows a snapshot of the instantaneous distribution of the oscillators on the limit cycle (from~\cite{Nakao}).}
        \label{fig-noise}
\end{figure}

For limit-cycle oscillators, Teramae and Tanaka~\cite{Teramae,Teramae2} showed, by using phase reduction theory, that this kind of synchronization occurs generally
when the oscillators are driven by common Gaussian noise (see also Goldobin and Pikovsky~\cite{Goldobin1}).
It has also been shown that common random impulses and other types of noise generally induce synchronization of limit-cycle oscillators~\cite{Arai},\cite{Pikovskii}-\cite{Kurebayashi}.
Figure~\ref{fig-noise}(a) shows synchronization of two non-interacting FitzHugh-Nagumo oscillators induced by common Gaussian white noise (simulation)~\cite{Nakao}, and
Fig.~\ref{fig-noise}(b) shows reproduction of the dynamics of a periodically flashing LED induced by the same sequence of random impulses
(experiment)~\cite{Arai}, respectively.

Suppose that a noise-driven limit-cycle oscillator is described by a stochastic dynamical equation
\begin{align}
	\frac{d}{dt} {\bf X}(t) = {\bf F}({\bf X}(t)) + \varepsilon \xi(t) {\bf e},
	\label{lcsde}
\end{align}
where $\xi(t)$ is stationary Gaussian noise.
For simplicity, we assume that the noise is applied only in the direction ${\bf e}$ in the state space, and that the correlation time of the noise is longer than the relaxation time of the oscillator state to $\chi$\footnote{There are some subtleties in the phase reduction of oscillators driven by white noise, and this assumption is used to avoid them. See Sec. 5 for a brief explanation.}.
The noise $\xi(t)$ is generally colored and satisfies
\begin{align}
\langle \xi(t) \rangle = 0, \quad 
\langle \xi(t) \xi(s) \rangle = C(t-s),
\end{align}
where $ \langle \cdots \rangle $ represents the statistical average
and $C(t)$ is the autocorrelation function of $\xi(t)$.
When the white-noise limit is taken and $\xi(t)$ is regarded as white Gaussian noise, we interpret Eq.~(\ref{phssde}) as a stochastic differential equation of Stratonovich type~\cite{Stochastic,Gardiner}.
By applying the phase reduction to Eq.~(\ref{lcsde}), we obtain a phase equation for the phase $\theta(t)$ of the oscillator as~\cite{Teramae,Teramae2}
\begin{align}
\frac{d}{dt}{\theta}(t) = \omega + \varepsilon Z(\theta(t)) \xi(t),
\label{phssde}
\end{align}
where $Z(\theta) = {\bf Z}(\theta) \cdot {\bf e}$ represents the phase sensitivity function of the oscillator with respect to perturbations applied in the direction ${\bf e}$.

We now consider two non-interacting oscillators with phase variables $\theta(t)$ and $\theta'(t)$ driven by the common noise $\xi(t)$, described by
\begin{align}
\frac{d}{dt}{\theta}(t) = \omega + \varepsilon Z(\theta(t)) \xi(t),
\quad
\frac{d}{dt}{\theta}'(t) = \omega + \varepsilon Z(\theta'(t)) \xi(t),
\label{twophssde}
\end{align}
and denote their phase difference as $\phi(t) = \theta'(t) - \theta(t)$.
To analyze whether these two oscillators can synchronize due to the common noise, we focus on the situation where $|\phi(t)|$ is sufficiently small, and characterize its growth (or decay) rate by the mean {\it Lyapunov exponent} 
\begin{align}
\Lambda = \left \langle \frac{d \ln |\phi(t)| }{ dt } \right \rangle,
\end{align}
which is averaged over the noise.
If $\Lambda < 0$, $|\phi(t)|$ shrinks on average and the oscillators will synchronize with each other.
If $\Lambda > 0$, $|\phi(t)|$ will grow exponentially and the oscillators will be desynchronized.

By subtracting the first equation from the second equation in Eq.~(\ref{twophssde}), we obtain the equation for the phase difference $\phi$ as
\begin{align}
	\frac{d}{dt}{\phi}(t) = \varepsilon \{ Z(\theta(t)+\phi(t)) - Z(\theta(t)) \} \xi(t),
\end{align}
and by expanding the phase sensitivity function as $Z(\theta + \phi) = Z(\theta) + Z'(\theta) \phi + O(\phi^{2})$, where $'$ indicates $d/d\theta$, we can obtain a linearized equation for the phase difference as
\begin{align}
\frac{d}{dt}{\phi}(t) = \varepsilon Z'(\theta(t)) \phi(t) \xi(t).
\end{align}
Therefore, the logarithm of the absolute phase difference $ \ln |\phi(t)| $ obeys 
\begin{align}
\frac{d}{dt} \ln |\phi(t)| = \varepsilon Z'(\theta(t)) \xi(t).
\label{phsdifsde}
\end{align}
Here, ordinary rules of differential calculus apply even if $\xi(t)$ is white, because the Stratonovich interpretation is used.

To calculate the mean Lyapunov exponent $\Lambda$, we need to average Eq.~(\ref{phsdifsde}) with respect to noise.
However, $\theta(t)$ and $\xi(t)$ on the right-hand side are not independent in general. They are correlated even if $\xi(t)$ is white, because the Stratonovich interpretation is used. For white noise, the equation can easily be analyzed by transforming it to a stochastic differential equation of Ito type~\cite{Arai,Stochastic,Gardiner}.
Here, we rather use the Novikov's theorem~\cite{Stochastic} to calculate this average, which states that the following formula holds for a general Gaussian stochastic process:
\begin{align}
\langle Z'(\theta(t)) \xi(t) \rangle = \int_0^t ds \langle \xi(t) \xi(s) \rangle \left\langle \frac{\delta Z'(\theta(t))}{\delta \xi(s)} \right\rangle.
\end{align}
Here, $\delta Z'(\theta(t)) / \delta \xi(s)$ denotes a functional derivative of $Z'(\theta(t))$ by $\xi(s)$.  This term can be calculated from 
\begin{align}
	\theta(t) = \theta(0) + \omega t + \varepsilon \int_0^t ds Z(\theta(s)) \xi(s),
\end{align}
which is a formal integration of Eq.~(\ref{phssde}), as
\begin{align}
\frac{\delta Z'(\theta(t))}{\delta \xi(s)}
=
\frac{d Z'(\theta(t))}{d \theta(t)}
\frac{\delta \theta(t)}{\delta \xi(s)}
=
\varepsilon Z''(\theta(t)) Z(\theta(s))
\end{align}
for $ s\leq t$. Using this result, we can average Eq.~(\ref{phsdifsde}) and obtain the mean Lyapunov exponent as
\begin{align}
\Lambda
=
\left \langle \frac{d}{dt} \ln |\phi(t)| \right \rangle
=
\varepsilon^2 \int_{0}^{t} ds C(t-s) \left\langle Z''(\theta(t)) Z(\theta(s)) \right\rangle,
\end{align}
where $ C(t-s) = \langle \xi(t) \xi(s) \rangle$.

As shown by Teramae and Tanaka~\cite{Teramae2}, the statistical average with respect to noise can be estimated as an average with respect to the probability density function $P(\theta(t), \theta(s)) = P(\theta(t)|\theta(s)) P(\theta)$, where $P(\theta)$ is the stationary probability density function of the phase $\theta$ and $P(\theta(t) |\theta(s))$ is the conditional probability density function of $\theta(t)$ given $\theta(s)$.
Because $\varepsilon$ is small, this can be approximated as $P(\theta) \simeq 1/2\pi$ and $P(\theta(t)|\theta(s)) \simeq \delta(\theta(t) - \theta(s) - \omega(t-s))$, where $\delta(t)$ represents Dirac's delta function. Namely, the oscillator phase almost uniformly spreads over $[0, 2\pi]$ statistically, and $\theta(t) \approx \theta(s) + \omega(t-s)$ holds, i.e., the oscillator phase approximately increases with the natural frequency $\omega$.
Thus, the Lyapunov exponent $\Lambda$ can be approximately calculated as
\begin{align}
\Lambda
&\simeq
\varepsilon^2 \int_{0}^{t} ds C(t-s) \frac{1}{2\pi} \int_0^{2\pi} d\theta Z''(\theta) Z(\theta+\omega(t-s)) \cr \cr
&\simeq 
\varepsilon^2 \int_{0}^{\infty} d\tau C(\tau) \frac{1}{2\pi} \int_0^{2\pi} d\theta Z''(\theta) Z(\theta-\omega \tau),
\end{align}
where we introduce $\tau = s - t$ and change the integration range to $0 \leq \tau < \infty$ in the second line, assuming that the autocorrelation function $C(\tau)$ decays quickly.

It can be shown~\cite{Teramae2}, using Fourier decomposition, that this $\Lambda$ is always nonpositive, i.e.,
\begin{align}
\Lambda = - \varepsilon^{2} \sum_{n=-\infty}^{\infty} n^{2} |Z_{n}|^{2} \frac{ \langle | \xi_{n \omega} |^{2} \rangle }{ 2 } \leq 0,
\end{align}
where $Z(\theta)=\sum_{n=-\infty}^{\infty} Z_{n} e^{i n \theta}$ and $\xi(t) = \int_{-\infty}^{\infty} \xi_{k} e^{i k t} dk$.
In particular, when $\xi(t)$ is Gaussian white noise of unit intensity, that is, when $C(t-s) = \langle \xi(t) \xi(s) \rangle = \delta(t-s)$, the Lyapunov exponent is simply given by
\begin{align}
\Lambda
=
\left \langle \frac{d}{dt} \ln |\phi(t)| \right \rangle 
=
\frac{\varepsilon^2}{2} \left\langle Z''(\theta(t)) Z(\theta(t)) \right\rangle,
\end{align}
where the factor of $1/2$ arises from the Stratonovich interpretation~\cite{Stochastic,Gardiner}.
In this case, by using partial integration and the $2\pi$-periodicity of $Z(\theta)$, it can easily be shown that
\begin{align}
\Lambda
\simeq
\frac{\varepsilon^2}{4\pi} \int_0^{2\pi} Z''(\theta) Z(\theta) d\theta
= - \frac{\varepsilon^2}{4\pi} \int_0^{2\pi} \left\{ Z'(\theta) \right\}^2 d\theta
\leq 0.
\end{align}

Thus, $\Lambda \leq 0$ generally holds under the phase reduction approximation.
The equality $\Lambda = 0$ holds only when the phase sensitivity function is strictly constant, i.e., $Z'(\theta) = 0$, which corresponds to an unphysical situation in which the oscillator phase does not respond to the perturbation at all.
Therefore, small phase differences between non-interacting oscillators driven by weak common noise will almost always shrink and the oscillators will eventually synchronize.
This phenomenon is called {\em noise-induced synchronization} or {\em stochastic synchronization} and has recently attracted considerable attention~\cite{Arai},\cite{Teramae}-\cite{Kurebayashi},\cite{Goldobin2}.
Figure~\ref{fig-noise}(c) shows the negativity of the Lyapunov exponent $\Lambda$ of the FitzHugh-Nagumo oscillator driven by weak Gaussian-white noise for $\varepsilon > 0$.

The above result can be extended to a more general class of noisy forcing by employing an effective white-noise approximation of the colored noise~\cite{Goldobin2}.
Moreover, as shown in Ref.~\cite{Nakao,Kurebayashi}, through analysis of the multidimensional Fokker-Planck equation corresponding to Eq.~(\ref{phssde}) with an additional independent noise term (under effective white-noise approximation if the noise is colored), not only $\Lambda$, but also the stationary probability distribution of the phase difference $\phi$ can be obtained as
\begin{align}
	P(\phi) \propto \frac{1}{D[q(0)-q(\phi)]+E q(0)}.
\end{align}
Here, $D$ and $E$ are the intensities of the common and independent noise, respectively, and $q(\phi) = (2\pi)^{-1} \int_{0}^{2\pi} Z(\psi) Z(\psi+\phi) d\psi$ is an autocorrelation function of the phase sensitivity function $Z(\theta)$.
Figure~\ref{fig-noise}(d) compares the above theoretical distribution of the phase difference $\phi$ with numerical simulations using $200$ non-interacting FitzHugh-Nagumo oscillators driven by common and independent noise terms~\cite{Nakao}. We can observe that the distribution of the phase difference $P(\phi)$ has a sharp peak at $\phi=0$, indicating synchronization of the oscillators induced by the common noise. Correspondingly, the snapshot of the oscillator states on the limit cycle shows a synchronized cluster.

Note that the above theories hold only for sufficiently small noise. If the noise intensity is too large, the phase reduction for weak perturbations cannot be applied.
Though we do not go into details in this article, oscillators driven by common random impulses can also be analyzed using phase reduction theory for impulsive perturbations~\cite{Arai,Nakao0}.
In this case, not only weak impulses but also strong impulses can be treated, provided that the intervals between the impulses are sufficiently large.
It has been shown both theoretically and experimentally that noise-induced synchronization occurs when the impulses are sufficiently weak; however, for stronger impulses, noise-induced desynchronization may also occur.
Synchronization of oscillators by common telegraph noise has also been analyzed similarly~\cite{Nagai}.

\section{Synchronization of coupled oscillators}

In this section, we analyze systems of weakly interacting limit-cycle oscillators using phase reduction theory~\cite{Kuramoto}. We first derive coupled phase equations for oscillators interacting weakly via general coupling networks. Then, we consider the mutual synchronization of a pair of coupled oscillators, the collective synchronization of globally coupled oscillator populations, and the dynamics of coupled oscillators on scale-free networks.

\subsection{Phase equation for coupled oscillators}

First, we derive coupled phase equations for a system of $N$ limit-cycle oscillators weakly interacting on a general network.
The dynamics of the $j$th oscillator state ${\bf X}_{j}$ ($j=1, \cdots, N$) are described by
\begin{align}
\frac{d}{dt}{\bf X}_j(t) = {\bf F}_j({\bf X}_j) + \varepsilon \sum_{k=1}^{N} {\bf G}_{jk}({\bf X}_j, {\bf X}_k),
\label{coupled}
\end{align}
where $ {\bf F}_j({\bf X}_j) $ represents the unperturbed dynamics of the oscillator, $ {\bf G}_{jk}({\bf X}_j, {\bf X}_k) $ represents the effect of oscillator $ k $ on oscillator $ j $, and $\varepsilon$ is a small parameter characterizing the intensity of the interaction.
It is assumed that the differences in the oscillator properties are small and $O(\varepsilon)$,
namely, the dynamics of each oscillator can be expressed as
\begin{align}
{\bf F}_j({\bf X}_j) = {\bf F}({\bf X}_j) + \varepsilon {\bf f}_j({\bf X}_j),
\end{align}
where $ {\bf F} $ is common to all oscillators and $ \varepsilon {\bf f}_j $ represents individual heterogeneity.
We assume that the common component $ d{\bf X}(t)/dt = {\bf F}({\bf X}) $ has a stable limit cycle $\chi$ of period $T$ and natural frequency $\omega = 2\pi / T$, and denote the oscillator state on $\chi$ as  $ {\bf X}_0(\theta) $.
The heterogeneity $\varepsilon {\bf f}_{j}$ will be treated as weak perturbations to the common component~\cite{Kuramoto}, together with the weak mutual interaction.
As previously, we denote the phase function of $\chi$ as $\Theta({\bf X})$ and the phase sensitivity function of $\chi$ as $ {\bf Z}(\theta) $.
The phase of the $j$th oscillator is given by $ \theta_j(t) = \Theta( {\bf X}_j(t) ) $.

Using phase reduction theory, we can simplify Eq.~(\ref{coupled}) describing coupled limit-cycle oscillators to coupled phase equations by treating the small heterogeneity of the oscillators and the mutual coupling as weak perturbations~\cite{Kuramoto}.
Similar to the derivation of Eq.~(\ref{phseq}), we approximate $ {\bf X}_j(t) $ in each term of Eq.~(\ref{coupled}) by $ {\bf X}_0(\theta_j(t)) $ on $\chi$.
The reduced approximate phase equations are given by
\begin{align}
\frac{d}{dt}{\theta}_j(t)
=
\omega +
\varepsilon {\bf Z}(\theta_j) \cdot \left[
{\bf f}_j(\theta_j) + \sum_{k=1}^{N} {\bf G}_{jk}(\theta_j, \theta_k)
\right],
\label{phs3}
\end{align}
where ${\bf f}_j(\theta_j) = {\bf f}({\bf X}_0(\theta_j)) $ and
$ {\bf G}_{jk}(\theta_j, \theta_k) = {\bf G}_{jk}({\bf X}_0(\theta_j), {\bf X}_0(\theta_k)) $.
As before, the error caused by this phase reduction is $O(\varepsilon^{2})$.

Further, we can perform the averaging approximation, assuming that the effect of
the mutual coupling and the heterogeneity of the oscillators are
sufficiently small.
To this end, we introduce new relative phase variables $ \phi_j(t) = \theta_j(t) - \omega t $ $(j=1, ..., N)$ by subtracting the steadily increasing component $\omega t$ from the
individual phase $ \theta_j(t) $ of the oscillators.
Equation~(\ref{phs3}) is then rewritten as
\begin{align}
\frac{d}{dt}{\phi}_j(t) =
 \varepsilon {\bf Z}(\phi_j + \omega t) \cdot \left[ {\bf f}_j(\phi_j + \omega t)
 + \sum_{k=1}^{N} {\bf G}_{jk}(\phi_j + \omega t, \phi_k + \omega t) \right],
\end{align}
which indicates that $\phi_j(t)$ is a slow variable, because the right-hand side is $ O(\varepsilon) $.
Thus, similar to the previous section, we may average the right-hand side over one period of the limit cycle, assuming that the relative phase variables $\phi_1(t), ..., \phi_{N}(t)$ do not vary in this period.
By this averaging approximation, we obtain a much simpler equation,
\begin{equation}
\frac{d}{dt} \phi_j(t) = \varepsilon \left[ \Delta_j + \sum_{k=1}^{N} \Gamma_{jk}(\phi_j - \phi_k) \right],
\end{equation}
for $j=1, ..., N$.
Here,
\begin{align}
\Delta_j = \frac{1}{2\pi} \int_{0}^{2\pi} {\bf Z}(\psi) \cdot {\bf f}_j(\psi) d\psi
\end{align}
gives the frequency deviation of the $j$th oscillator from $\omega$, and
\begin{align}
\Gamma_{jk}(\varphi)
=
 \frac{1}{2\pi} \int_{0}^{2\pi} {\bf Z}(\varphi + \psi) \cdot {\bf G}_{jk}(\varphi + \psi,\ \psi) d\psi
\end{align}
is called the {\em phase coupling function}~\cite{Kuramoto}, which represents the effect of oscillator $ k $ on oscillator $ j $ over one period of the limit-cycle oscillation.
As before, the error of this averaging approximation is generally $O(\varepsilon^{2})$.

Returning to the original phase variables, the averaged equation is given by
\begin{equation}
\frac{d}{dt} \theta_j(t) = \omega_j + \varepsilon
\sum_{k=1}^{N} \Gamma_{jk}(\theta_j - \theta_k)
\label{cplphseq}
\end{equation}
for $j=1, ..., N$, where $ \omega_j = \omega + \varepsilon \Delta_j $ gives the natural frequency of $j$th oscillator\footnote{As previously, the phase $\tilde{\theta}_{j}$ after averaging is slightly different from the raw phase $\theta_{j}$ before averaging, but they are treated as identical here. When the coupling is absent, $\tilde{\theta}_{j}$ increases with a constant $\omega_{j}$, while $\theta_{j}$ may exhibit additional tiny oscillations due to the effect of the heterogeneity ${\bf f}_{j}$.}.
This gives a general form of the coupled phase model on a network.
It is important that the phase coupling function $\Gamma_{jk}$ in the averaged equation depends only on the phase difference $ \theta_j - \theta_k $, because this simplifies the analysis of the synchronized state significantly.

\subsection{A pair of symmetrically coupled oscillators}

Let us analyze the simplest case of two symmetrically coupled oscillators using the derived phase model~(\ref{cplphseq}).
It is often mentioned in the literature that, in 17th century, Dutch physicist C. Huygens noticed that two pendulum clocks on a wall synchronize with each other as a result of weak interaction through the wall~\cite{Pikovsky}. This observation can easily be reproduced using two metronomes placed on a movable plate, as demonstrated by many researchers recently (see YouTube for videos!).
In a recent interesting experiment, Aihara~\cite{Aihara} found that the calls of
two frogs caught in a rice field exhibit anti-phase synchronization, and explained
this phenomenon using the phase model.

Let us consider a pair of oscillators, denoted as $1$ and $2$, weakly interacting via a symmetric coupling function ${\bf G}$ as
\begin{align}
\frac{d}{dt}{\bf X}_1(t) = {\bf F}_1({\bf X}_1) + \varepsilon {\bf G}({\bf X}_1, {\bf X}_2),
\quad
\frac{d}{dt}{\bf X}_2(t) = {\bf F}_2({\bf X}_2) + \varepsilon {\bf G}({\bf X}_2, {\bf X}_1).
\end{align}
We assume that the oscillators are similar to each other, i.e., the difference between ${\bf F}_{1}({\bf X})$ and ${\bf F}_{2}({\bf X})$ is $O(\varepsilon)$.
Then, from Eq.~(\ref{cplphseq}), these equations can be reduced to the following coupled phase equations:
\begin{align}
\frac{d}{dt}{\theta}_1(t) = \omega_1 + \varepsilon \Gamma(\theta_1 - \theta_2),\;\;\;
\frac{d}{dt}{\theta}_2(t) = \omega_2 + \varepsilon \Gamma(\theta_2 - \theta_1),
\label{twocplphs}
\end{align}
where $\omega_{1}$ and $\omega_{2}$ are the natural frequencies of oscillators $1$ and $2$, respectively, and the phase coupling function is given by
\begin{align}
\Gamma(\varphi) = \frac{1}{2\pi} \int_{0}^{2\pi} {\bf Z}(\varphi + \psi) \cdot {\bf G}({\bf X}_{0}(\varphi + \psi),\ {\bf X}_{0}(\psi)) d\psi.
\end{align}

To analyze whether these two oscillators synchronize or not, we consider the phase difference $ \varphi(t) = \theta_1(t) - \theta_2(t) $ between them. From Eq.~(\ref{twocplphs}), $\varphi$ obeys
\begin{align}
\frac{d}{dt}{\varphi}(t) = \varepsilon \left\{ \Delta + \Gamma_{a} ( \varphi ) \right\},
\label{2oscphsdif}
\end{align}
where $ \varepsilon \Delta = \omega_1 - \omega_2 $ is the frequency difference between the two oscillators and
\begin{align}
	\Gamma_{a} ( \varphi  ) = \Gamma( \varphi  ) - \Gamma( -\varphi  )
\end{align}
represents the {\em antisymmetric} part of the phase coupling function (multiplied by two).
It is clear that $ \Gamma_{a}(0) = \Gamma_{a}(\pm \pi) = 0 $ holds, because
$ \Gamma( \varphi ) $ is $ 2\pi $-periodic.
The above equation~(\ref{2oscphsdif}) for a pair of coupled oscillators is similar to Eq.~(\ref{phsperiodsingle}) for the oscillator driven by periodic forcing, and can therefore be analyzed in a similar manner.
Specifically, if Eq.~(\ref{2oscphsdif}) has a stable fixed point at which $d\varphi(t)/dt = 0$, the oscillators $1$ and $2$ can mutually synchronize with a constant phase difference determined by the stable fixed point\footnote{Again, the two oscillators synchronize within the averaging approximation; when observed using the phase variables before averaging, the phase difference does not grow but exhibits tiny oscillations around the phase-locking point.}.

If the two oscillators have equal frequency, i.e., $ \omega_1 = \omega_2 $  ( $ \Delta = 0 $ ), Eq.~(\ref{2oscphsdif}) always has an {\em in-phase} fixed point with $ \varphi = 0 $ and an {\em anti-phase} fixed point with $ \varphi = \pm \pi $.
The in-phase fixed point is stable if the slope $ \Gamma_{a}'(0) $ of $ \Gamma_{a}( \varphi ) $ at the origin is negative, i.e., $ \Gamma_{a}'(0) < 0$.
Therefore, the two oscillators can exhibit in-phase synchronization; the phase coupling function is said to be {\em attractive} in this case.
In contrast, if $\Gamma_{a}'(0) > 0$, the in-phase state is unstable.
Similarly, the anti-phase state with $\varphi = \pi$ is stable when $ \Gamma_{a}'(\pi)  < 0$ and unstable when $ \Gamma_{a}'(\pi) > 0$.
Depending on the functional form of $ \Gamma_{a}(\varphi) $, other fixed points may also exist.

If $ \Delta \neq 0 $, Eq.~(\ref{2oscphsdif}) has (at least) a pair of stable and unstable fixed points as long as
\begin{align}
\mbox{min}\ \Gamma_{a}(\varphi) < -\Delta < \mbox{max}\ \Gamma_{a}(\varphi)
\end{align}
is satisfied.
Then, the phase difference $ \varphi(t) $ converges to one of the stable fixed points and mutual synchronization (phase locking) occurs there. In this case, the stable phase differences are not simply given by $\varphi=0$ and $\varphi=\pi$ in general.
Finally, when $ |\Delta| $ is too large, Eq.~(\ref{2oscphsdif}) does not have any fixed points and $ \varphi(t) $ continues to increase or decrease.

Figure~\ref{rdsync}(a) shows an example of the function $\Gamma_{a}(\varphi)$, where both in-phase and anti-phase synchronized states are stable because $\Gamma_{a}'(0)$ and $\Gamma_{a}'(\pm \pi)$ are both negative. There exist two other synchronized states with some intermediate phase differences, which are both unstable because the slopes of $\Gamma_{a}(\varphi)$ are positive there. Figure~\ref{rdsync}(b) exhibits typical time sequences of the two oscillators in the in-phase and anti-phase synchronized states.
Actually, these figures are for two coupled reaction-diffusion systems rather than for two coupled low-dimensional oscillators, but the phase equation after reduction is the same as for ODEs.

\subsection{Globally coupled oscillators}

As the next example, we consider a population of $N$ oscillators interacting with all other oscillators. The dynamics of the $j$th oscillator is described by
\begin{align}
	\frac{d}{dt}{\bf X}_j(t) = {\bf F}_j({\bf X}_j) + \varepsilon \frac{1}{N} \sum_{k=1}^{N} {\bf H}({\bf X}_{k})
\label{gbcpl}
\end{align}
for $j=1, ..., N$, where the last term is the weak global coupling and ${\bf H}({\bf X}_{k})$ represents the effect of the $k$th oscillator.
Thus, all oscillators feel the common mean field $\sum_{k} {\bf H}({\bf X}_{k}) / N$, which is generated by the oscillators themselves.
This type of interaction is often called ``global'', ``mean-field'', or ``all-to-all'' coupling.
Such coupling arises when the interaction between the oscillators is sufficiently fast and uniformly works irrespective of their distance.
For example, a series array of the Josephson junctions has been considered
a typical example of globally coupled oscillators and intensively studied~\cite{Wiesenfeld}.
Other representative examples are the electrochemical oscillators coupled by a common resistor~\cite{Hudson} and the population of oscillatory catalytic particles in well-stirred chemical solution~\cite{Taylor}, for which collective synchronization transition has been clearly observed.
Population of rhythmically walking pedestrians on a bridge or population of fireflies interacting via flashing light may also be considered as globally coupled oscillators~\cite{Strogatz0,Strogatz2}.
Global coupling is also conceptually important as the simplest approximation of more complex interaction networks.

\begin{figure}[bt]
        \centering
       	\includegraphics[width=\hsize,clip]{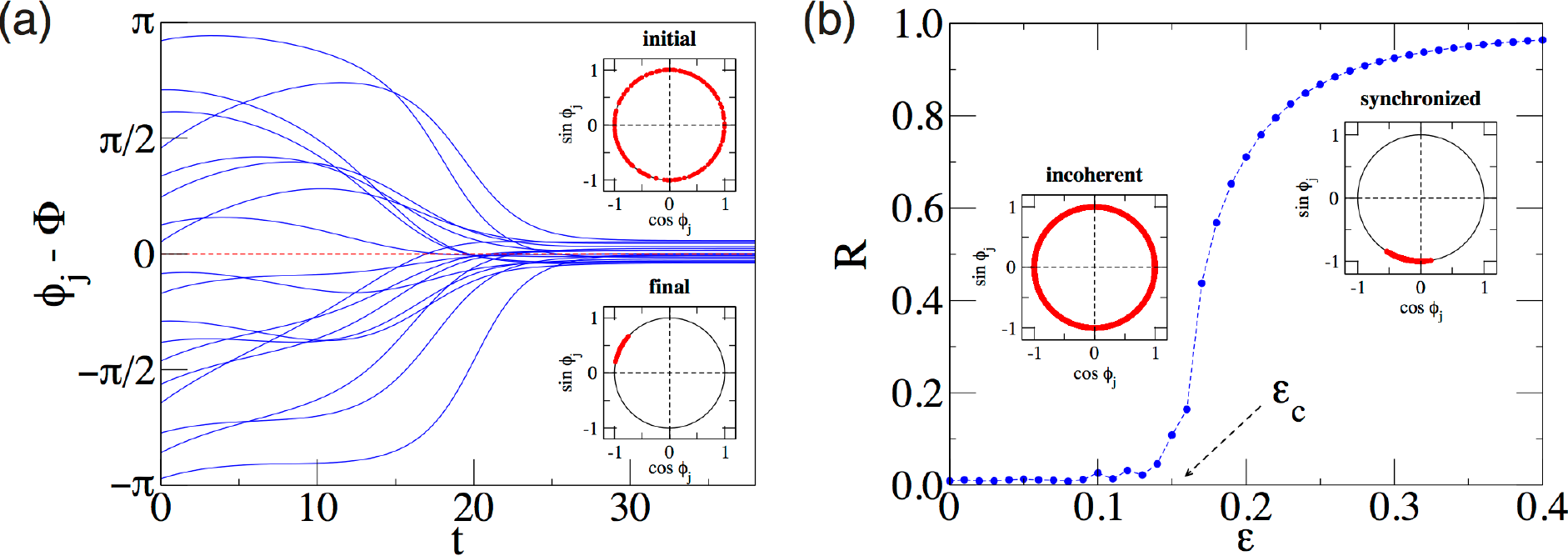}
        \caption{Collective synchronization of the Kuramoto model. (a) Dynamics of the relative phase variables of the oscillators ($N=100$, only $18$ oscillators are shown). (b) Order parameter $R$ vs. the coupling intensity $\varepsilon$ ($N=8000$). The inset shows oscillator distributions in the incoherent (left) and synchronized (right) states.
        }
        \label{fig-kuramoto}
\end{figure}

By applying the phase reduction and averaging approximation to Eq.~(\ref{gbcpl}) with ${\bf G}_{jk}({\bf X}_j, {\bf X}_k) = {\bf H}({\bf X}_k)/N$, the following {\em globally coupled phase model is} obtained:
\begin{align}
\frac{d}{dt}{\theta}_j(t) = \omega_j + \varepsilon \frac{1}{N} \sum_{k=1}^{N} \Gamma(\theta_j - \theta_k)
\end{align}
for $j=1, ..., N$, where $\omega_{j}$ is the natural frequency of the $j$th oscillator and
\begin{align}
\Gamma(\varphi) = \frac{1}{2\pi} \int_{0}^{2\pi} {\bf Z}(\varphi + \psi) \cdot {\bf H}({\bf X}_{0}(\psi)) d\psi
\end{align}
is the phase coupling function.

In particular, when the phase coupling function is given by the lowest order Fourier term as $ \Gamma(\varphi) = - \sin \varphi$, the model becomes the well-known {\em Kuramoto model}~\cite{Kuramoto,Acebron,Strogatz4}, given by
\begin{align}
	\frac{d}{dt}{\theta}_j(t) = \omega_j - \varepsilon \frac{1}{N} \sum_{k=1}^{N} \sin(\theta_j - \theta_k)
\label{kuramoto}
\end{align}
for $j=1, ..., N$.
This model describes a system of globally coupled phase oscillators with attractive coupling when $\varepsilon > 0$.
It is usually assumed that $\omega_{j}$ is randomly drawn from a probability density function $g(\omega)$.
The typical setting is that $g(\omega)$ is a symmetric one-humped smooth distribution, e.g., Gaussian or Lorentzian~\cite{Kuramoto}.
When the phase coupling function is given in the form $ \Gamma(\varphi) = - \sin(\varphi + \alpha)$, with the parameter $\alpha$ representing the coupling phase shift, the corresponding phase model is often called the {\em Sakaguchi-Kuramoto model}~\cite{Sakaguchi}.
Note that the coupling is attractive when $|\alpha| < \pi/2$, because $\Gamma_a(\varphi) = - 2 \sin \varphi \cos \alpha$.
Such sinusoidal coupling function is rigorously obtained for the case of coupled Stuart-Landau oscillators (see Appendix~\ref{App:SL}).
More general coupling functions with higher-order harmonics have also been considered in detail by Crawford~\cite{Strogatz4,Crawford}, Daido~\cite{Daido1}, and others.

As shown by Kuramoto, Eq.~(\ref{kuramoto}) exhibits {\em collective synchronization transition}.
In this model, the attractive coupling tends to synchronize the oscillators, while the dispersion in $\omega_{j}$ tends to suppress synchronization.
As a result of these competing effects, the oscillators, which are incoherent when $\varepsilon$ is sufficiently small, suddenly begin to synchronize when $\varepsilon$ exceeds a certain critical value $\varepsilon_{c}$.
Figure~\ref{fig-kuramoto}(a) shows the dynamics of relative phase variables (see below) for $\varepsilon > \varepsilon_{c}$ undergoing mutual synchronization, and (b) shows the sudden increase in the degree of synchronization as a function of $\varepsilon$.
This sudden transition is now called the Kuramoto transition\footnote{Collective synchronization of oscillators was numerically observed by Winfree already in 1967~\cite{Winfree1} for globally coupled phase oscillators before averaging, $d\theta_{j}/dt = \omega_{j} + \varepsilon Z(\theta_{j}) \sum_{k} X(\theta_{k})$, where $X(\theta_{k})$ is the waveform of the $k$th oscillator.}.
Collective synchronization transition of coupled oscillators has been experimentally observed in a system of coupled electrochemical oscillators~\cite{Hudson}, and in an ensemble of porous catalytic particles undergoing Belousov-Zhabotinsky chemical oscillations~\cite{Taylor}.
The wobbling of the millennium bridge in London, which was caused by the collective synchronization of many pedestrians, is also considered a real-world example of this phenomenon~\cite{Strogatz2}.

The degree of collective synchronization can be characterized by the {\em complex order parameter} or the {\em mean field}~\cite{Kuramoto}, 
\begin{align}
R(t) \exp( i \Phi(t) ) = \lim_{N \to \infty} \frac{1}{N} \sum_{k=1}^{N} \exp( i\theta_k(t) ),
\label{order}
\end{align}
whose modulus $R$ and argument $\Phi$ quantify the amplitude and phase of the collective oscillations, respectively.
If the oscillators are synchronized and steady collective oscillations are generated, $R$ takes a non-zero constant value within $0 < R \leq 1$, where $R=1$ corresponds to the complete synchronization of all oscillators, and the collective phase $\Phi$ increases with a constant collective $\Omega$ as $\Phi(t) = \Phi(0) + \Omega t$ in the $N \to \infty$ limit.
The collective frequency $\Omega$ may differ from the simple mean of $\omega$ with respect to the frequency distribution $g(\omega)$, depending on the functional forms of $g(\omega)$ and $\Gamma(\varphi)$.
When the oscillators are not at all synchronized and behave independently (which is referred to as being {\em incoherent}), the amplitude is $R=0$ and $\Phi$ is not defined.
For finite $N$, $R$ and $\Phi$ generally exhibit finite-size fluctuations of $O(1/\sqrt{N})$.

The Kuramoto model given by Eq.~(\ref{kuramoto}) can be analytically solved in the $N \to \infty$ limit when $g(\omega)$ is chosen appropriately.
This is because the Kuramoto model with the sinusoidal phase coupling function can be formally rewritten, using $R(t)$ and $\Phi(t)$, in the form of a single oscillator driven by periodic external forcing, as
\begin{align}
\frac{d}{dt}{\theta}_j(t) = \omega_j - \varepsilon R(t) \sin(\theta_j - \Phi(t))
\label{kura2}
\end{align}
for $j=1, ..., N$.
Indeed, by assuming that steady collective oscillations are formed with $R(t) \equiv R_s = const.$, $\Omega \equiv \Omega_{s} = const.$, and $\Phi(t) = \Omega_{s} t$, where $R_{s}$ and $\Omega_{s}$ are the steady amplitude and frequency of the collective oscillations, and introducing relative phase variables $\phi_{j}(t) = \theta_{j}(t) - \Omega_{s} t$, Eq.~(\ref{kura2}) can be cast into the form
\begin{align}
\frac{d}{dt}{\phi}_j(t) = \omega_j - \Omega_{s} - \varepsilon R_{s} \sin \phi_{j},
\end{align}
which is similar to Eq.~(\ref{phsperiodsingle}) for a periodically driven oscillator and can therefore be analyzed similarly (see~\cite{Kuramoto} for the details).
The oscillators are separated into two groups depending on their natural frequencies $\omega_{j}$, namely, the synchronous group that is phase-locked to the mean field, and the asynchronous group that undergoes phase slips.

However, there is one essential difference between this case and that of the oscillator driven by external forcing -- {\em the mean field is spontaneously produced by the oscillators themselves} in the present situation.
That is, if the values of $R_{s}$ and $\Omega_{s}$ are given, the dynamics of a single oscillator is determined from Eq.~(\ref{kura2}).
In turn, these driven oscillators produce the mean field expressed in Eq.~(\ref{order}), which we denote by $R_{s}'$ and $\Omega_{s}'$ to distinguish from the given $R_{s}$ and $\Omega_{s}$.
Thus, the given values of $R_{s}$ and $\Omega_{s}$ should coincide with the resulting values of $R_{s}'$ and $\Omega_{s}'$. This type of argument is called the {\it self-consistency theory}, and is well known in statistical physics.
From such analysis, the self-consistency equation for $R_{s}$ is obtained in the form $R_{s} = h(\varepsilon R_{s})$, where $h(x)$ is a nonlinear function of $x$, and, from this equation, the critical value $\varepsilon_{c}$ and the dependence of $R_{s}$ on $\varepsilon$ near the transition point are obtained~\cite{Kuramoto}.
In the case of the Kuramoto model with symmetric, smooth, and one-humped $g(\omega)$, the order parameter increases as $R_{s} \sim (\varepsilon - \varepsilon_{c})^{1/2}$ for $\varepsilon > \varepsilon_{c}$ with a critical exponent $1/2$. If the phase coupling function $\Gamma(\varphi)$ has higher-harmonic components, this exponent can be different~\cite{Daido1}.
Moreover, if $g(\omega)$ has a compact support, the synchronization transition can even be discontinuous~\cite{Pazo}.

In a recent study, Ott and Antonsen showed that, for the Kuramoto model with Lorentzian frequency distribution, the whole infinite-dimensional system can be drastically reduced to a single ODE for the complex order parameter by putting an ansatz (called the Ott-Antonsen ansatz) for the probability density function of the oscillator phase, from which the exponent $1/2$ immediately follows~\cite{OttAntonsen,Martens}.
The case with bimodal $g(\omega)$ has also been analyzed using this ansatz~\cite{Martens}.
Moreover, Chiba performed a mathematically rigorous analysis of the Kuramoto transition based on the Gelfand triple~\cite{Chiba}. The behavior of the order parameter near the transition was rigorously determined by the center-manifold reduction of the system near the transition point.
See also~\cite{PikovskyKurRev} for recent developments in the analysis of globally coupled oscillators.

Finally, although we do not discuss here, the following model of globally
coupled identical oscillators with {\em additive} noise has also been well studied:
\begin{align}
	\frac{d}{dt}{\theta}_j(t) = \omega + \varepsilon \frac{1}{N} \sum_{k=1}^{N} \Gamma(\theta_j - \theta_k) + \xi_{j}(t),
\end{align}
where $\xi_{j}(t)$ is the Gaussian white noise applied independently to each oscillator. This model also exhibits collective synchronization transition and can be analyzed in the framework of the nonlinear Fokker-Planck equation~\cite{Kuramoto}.
The critical exponent of the order parameter is generally $1/2$ for this system.
See~\cite{Kuramoto,Acebron,Strogatz4} for the detailed analysis of the Kuramoto and related models.

\subsection{Coupled oscillators on complex networks}

Here, we consider coupled phase oscillators on complex networks as a relatively recent topic of interest.
Complex network structures, which are typically characterized by their small-world and scale-free properties, are now regarded as universal understructures in the real world.
The small-world property emphasized by Watts and Strogatz~\cite{Strogatz3} and the scale-free property emphasized by Barab\'asi and Albert~\cite{Barabasi} have been observed in various real-world systems, ranging from biochemical reaction networks in cells to computer networks.
Since their pioneering studies, coupled dynamical systems interacting via complex networks have attracted considerable attention. In particular, Kuramoto-type models of coupled oscillators on complex networks have been studied extensively~\cite{Ichinomiya}-\cite{Nakao3}.

Let us consider a system of $N$ oscillators, $j=1, \cdots, N$, that are weakly coupled via a network,
\begin{align}
\frac{d}{dt}{\bf X}_j(t) = {\bf F}_{j}({\bf X}_j) + \varepsilon \sum_{k=1}^{N} A_{jk} {\bf G}({\bf X}_j, {\bf X}_k),
\end{align}
where the network topology is specified by the $N \times N$ adjacency matrix $A_{jk}$.
Each component of the adjacency matrix takes $A_{jk}=1$ when there is a connection between oscillators $j$ and $k$, and $A_{jk}=0$ otherwise. For simplicity, we assume that the network is connected (there is a path between arbitrary oscillators), the coupling is symmetric ($ A_{jk} = A_{kj}$), and the coupling function ${\bf G}$ between the oscillators is identical.
In the following, we use the Barab\'asi-Albert scale-free random network~\cite{Barabasi} as the coupling network, which is characterized by a power-law distribution of the degrees (number of connections). 
See Fig.7~(a) for a typical realization of the network, where the oscillators with larger degrees are plotted in the center and  oscillators with smaller degrees are plotted in the periphery.

\begin{figure}[bt]
	\centering
	\includegraphics[width=\hsize,clip]{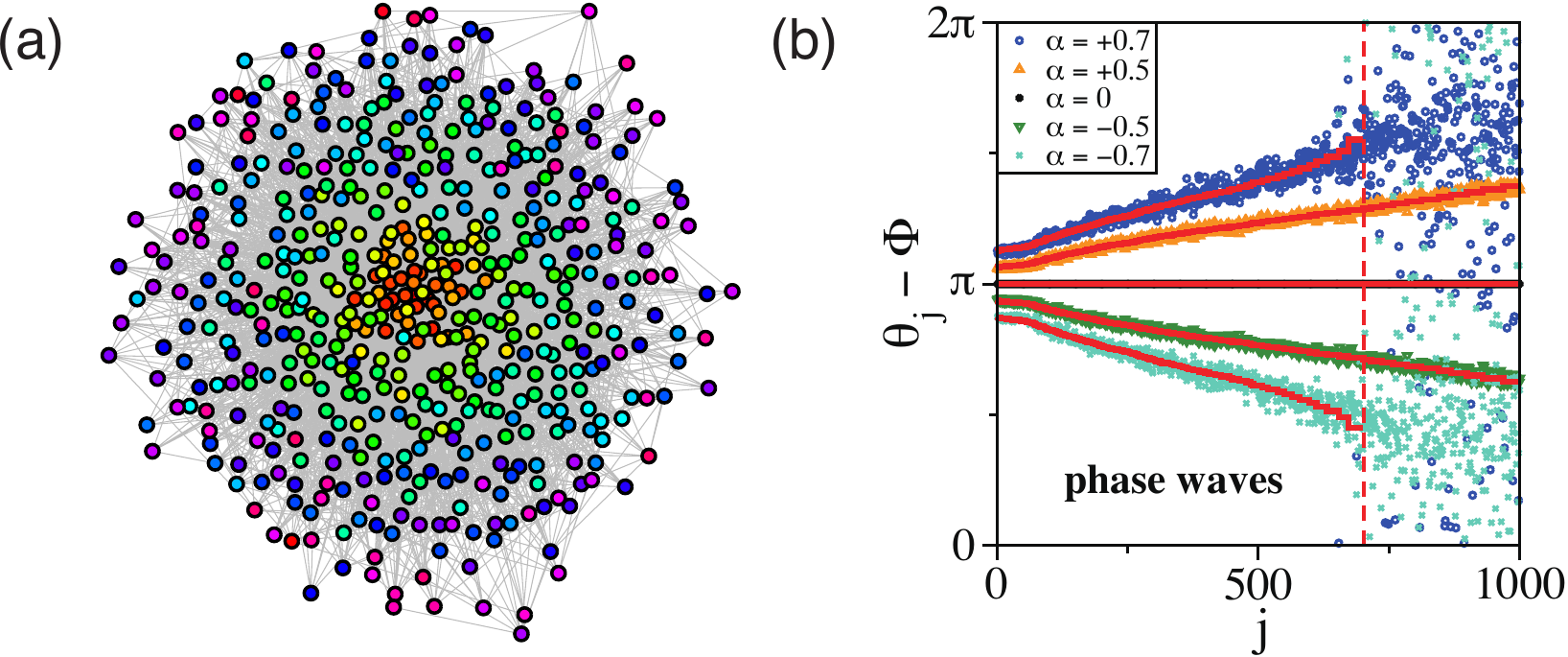}
	\caption{(a) Snapshot of the phase wave on a Barab\'asi-Albert scale-free network of size $N=1000$ and mean degree $20$ (only $500$ oscillators are shown), where the color of each oscillator represents its phase value. The parameters are $\omega_{j} \equiv 1 \; (j=1, ..., N)$, $\varepsilon = 0.5$, and $\alpha = -0.5$. (b) Snapshots of the phase waves plotted as functions of the oscillator index $j$ for several values of $\alpha$. The oscillators are sorted in decreasing order of degree $ n_{j} $. The solid curves represent the results of the self-consistent mean-field approximation.}
	\label{fig-network}
\end{figure}

From Eq.~(\ref{cplphseq}), the reduced phase equation for the $j$th oscillator ($j=1, \cdots, N$) is given by
\begin{align}
\frac{d}{dt}{\theta}_j(t) = \omega_{j} + \varepsilon \sum_{k=1}^{N} A_{jk} \Gamma(\theta_j - \theta_k),
\label{network}
\end{align}
where $\omega_{j}$ is the natural frequency and
\begin{align}
\Gamma(\varphi) = \frac{1}{2\pi} \int_{0}^{2\pi} {\bf Z}(\varphi + \psi) \cdot {\bf G}({\bf X}_{0}(\varphi + \psi), {\bf X}_{0}(\psi)) d\psi
\end{align}
is the phase coupling function.
We here assume a simple sinusoidal phase coupling function of Sakaguchi-Kuramoto type,
\begin{align}
 \Gamma(\varphi) = - \sin (\varphi + \alpha),
\label{sakakura}
\end{align}
where $\alpha$ is the phase shift of the coupling. This coupling is assumed to be attractive, i.e., $|\alpha| < \pi/2$ is satisfied~\cite{Kuramoto,Sakaguchi}.

Figure~\ref{fig-network}(a) shows a typical snapshot of the oscillators in the steady synchronized state after initial transition, where the color of each oscillator represents its phase value.
The natural frequencies of the oscillators are assumed identical in this case.
In Fig.~\ref{fig-network}(b), the relative phase $\theta_{j} - \Phi$ of the oscillator from the collective phase $\Phi$ is plotted as a function of the oscillator index $j$ for several values of $\alpha$ (the solid lines are the predictions of the self-consistency theory).
Here, the indices of the oscillators are sorted in decreasing order of the degrees, so small values of $j$ represent the high-degree oscillators and large values of $j$ the low-degree oscillators.
We can observe formation of heterogeneous steady phase waves that appear to propagate from high-degree oscillators at the center to low-degree oscillators near the periphery, or vice versa, whose slopes depend on $\alpha$. 
The oscillators are mutually synchronized when $|\alpha|$ is small, but low-degree oscillators start to desynchronize when $|\alpha|$ becomes larger.
These apparent phase waves are actually not real waves, but they are resulting from phase-locking dynamics of the oscillators to the mean field with gradually varying phase shifts,
as we explain below.

Although exact analysis of Eq.~(\ref{network}) is difficult for general random networks, we can understand its dynamics by employing the {\em mean-field approximation} of the scale-free network~\cite{Vespignani}.
Specifically, we ignore the actual couplings between the oscillators and assume that each oscillator simply feels the mean field over the network with an intensity proportional to its degree.
To this end, we define the ``local field'' experienced by oscillator $j$ as
\begin{align}
	R_j(t) \exp( i \Phi_j(t) ) = \sum_{k=1}^{N} A_{jk} \exp( i \theta_k(t) ).
\end{align}
Using this quantity, the dynamics of the $j$th oscillator can be rewritten as
\begin{align}
	\frac{d}{dt}{\theta}_j(t) = \omega_{j} - \varepsilon R_j(t) \sin(\theta_j - \Phi_j(t) + \alpha)
\end{align}
for $j=1, ..., N$, in a similar manner to the case of the Kuramoto model with global coupling.
This equation is still equivalent to Eq.~(\ref{network}) and, therefore, cannot be solved.

To proceed, we adopt the mean-field approximation, that is, we ignore the detailed topology of the network and retain the degrees of the oscillators $ n_{j} = \sum_{k=1}^{N} A_{jk} $ only.
We introduce a degree-weighted mean field over the entire network, such that
\begin{align}
	R(t) \exp(i \Phi(t)) = \sum_{k=1}^{N} \frac{n_k}{n_{total}} \exp(i \theta_k(t)),
	\;\;\;\;
	n_{total} = \sum_{k=1}^{N} n_k,
\end{align}
and approximate the amplitude $ R_j(t) $ and phase $ \Phi_j(t) $ of the local field as 
\begin{align}
	R_j(t) \simeq n_j R(t), \;\;\; \Phi_j(t) \simeq \Phi(t).
\end{align}
Namely, each oscillator is affected by and contributes to the mean field with the weight proportional to its degree $n_{j}$.
It is known that this kind of crude approximation works rather well for random networks with small diameters, in particular for scale-free networks
~\cite{Ichinomiya}-\cite{Vespignani}. 
Under this approximation, the phase equation for the $j$th oscillator is given by
\begin{align}
\frac{d}{dt}{\theta}_j(t) = \omega_{j} - \varepsilon n_{j} R(t) \sin( \theta_j - \Phi(t) + \alpha ).
\end{align}
Thus, each oscillator formally obeys the equation for a single oscillator driven by external forcing, characterized by the amplitude $R(t)$ and the collective phase $\Phi(t)$.
The above equation is similar to Eq.~(\ref{kura2}) for the Kuramoto model, with global coupling, and can therefore be analyzed in a similar, self-consistent way.
One essential difference from the Kuramoto model with global coupling is that the effective coupling intensity $ \varepsilon n_{j} $ is proportional to $ n_{j} $,
which can lead to heterogeneous dynamics of the oscillators on the network when $\alpha \neq 0$,
as shown in Fig.~\ref{fig-network}.

Using the above type of mean-field approximation, Ichinomiya~\cite{Ichinomiya} analyzed the Kuramoto model ($\alpha=0$) on a scale-free network and showed that $\varepsilon_{c}$ approaches $+0$ in the $N \to \infty$ limit; therefore, collective synchronization transition occurs even when $\varepsilon$ is vanishingly small (see also Restrepo, Ott, and Hunt~\cite{Restrepo}).
This result has attracted considerable attention, because it is analogous to the well-known result obtained by Pastor-Satorras and Vespignani for the epidemics model on scale-free networks~\cite{Vespignani}, which shows that epidemics can prevail on scale-free networks even if the infection rate is vanishingly small.
Ko and Ermentrout~\cite{Ko} also analyzed the identical frequency case (i.e., $\omega_{j} \equiv \omega$) self-consistently, and showed that heterogeneous steady states
consisting of phase-locked and phase-slipping oscillators, similar to those shown in Fig.~\ref{fig-network}, can be formed. Their result is interesting, because it shows the possibility of self-organized dynamical order on a network that is not simply completely synchronous.

\subsection{Other coupling schemes}
Though we do not discuss in this article, a wide variety of coupling schemes between the oscillators have been considered in the literature.
The local or diffusive coupling, where each oscillator interacts only with its nearest neighbors, is the most fundamental coupling scheme and describes, for example, spatially extended oscillatory chemical media such as the Belousov-Zhabotinsky reaction. Diffusively coupled phase models and their continuum limit have been extensively studied, and shown to exhibit propagating phase waves, target waves, spatiotemporal chaos, etc.~\cite{Kuramoto,Ermentrout1}.
Recently, nonlocally coupled phase oscillators, where the interaction between the oscillators decays with their distance, have also attracted attention. In particular, the ``chimera states'' in nonlocally coupled phase models have been intensively studied~\cite{Chimera1,Chimera2}. In the chimera state, the oscillators in the system split into coherent synchronized groups and incoherent desynchronized groups even if their natural frequencies are identical, and this coexisting steady state is maintained for a long time.
Phase models with time-delayed interactions have also been studied extensively~\cite{Ermentrout1,Schuster,Yeung,Ermentrout3}, because the effect of interaction delays are not negligible in many real-world systems such as coupled lasers or spiking neurons, and interesting dynamical effects, which cannot be observed without delays, have been reported.
There are numerous studies on various kinds of coupled-oscillator models and their intriguing dynamics. Interested readers are advised to look into the literature.

\section{Some recent topics on phase reduction theory}

In this last section, we comment on some of the recent topics and advances in phase reduction theory. We discuss the subtleties of phase equations for noisy oscillators, the application of phase reduction theory to rhythmic patterns in spatially extended systems, and a few other recent topics.

\subsection{Phase reduction of noisy oscillators}

In the analysis of the noise-induced synchronization (Sec. 4), it was assumed that the correlation time of the noise is longer than the relaxation time of the oscillator state to return to the limit cycle $\chi$, in order to avoid subtleties in the phase reduction. This issue is interesting, because it is a typical situation that arises in physics problems with two small timescales~\cite{Stochastic,Gardiner}; therefore, we comment on it briefly here.

In the analysis by Teramae and Tanaka~\cite{Teramae}, it was assumed that the reduced phase equation takes the conventional form, i.e., Eq.~(\ref{phssde}), even if the noise $\xi(t)$ is white Gaussian.
However, Yoshimura and Arai~\cite{Yoshimura} later pointed out that, when the noise is white, the following reduced phase equation should be used instead of the conventional one:
\begin{align}
	\frac{d}{dt}\theta(t) = \left[ \omega + \varepsilon^{2} Y(\theta) \right]
  + \varepsilon Z(\theta) \xi(t),
  \label{yoshimu}
\end{align}
which has an additional term $\varepsilon^{2} Y(\theta)$. This term arises from the relaxation dynamics of the oscillator state to $\chi$; see~\cite{Yoshimura} for the actual expression for $Y(\theta)$.

Here, one should remind that the white noise is also an approximation of real physical noise that has a small correlation time. Thus, there exist two small timescales in this problem: the correlation time $\tau_{\xi}$ of the noise and the relaxation time $\tau_{\rho}$ of the oscillator state (for simplicity, we assume that $\tau_{\rho}$ does not depend on $\theta$ here).
The non-agreement between the two equations is due to the order in which the white-noise limit ($\tau_{\xi} \to 0$) and the ``phase limit'' ($\tau_{\rho} \to 0$) are taken.
The conventional phase equation is obtained by taking the phase limit first and then taking the white-noise limit, while the additional term appears if the white-noise limit is first calculated, followed by the phase limit.
This problem was solved in a general manner in Ref.~\cite{Teramae3}, where it was shown through the application of singular perturbation theory that a family of effective white-noise phase equations of the form
\begin{align}
	\frac{d}{dt}\theta(t) = \left[ \omega + \frac{\varepsilon^{2} }{1 + (\tau_{\xi} / \tau_{\rho})} Y(\theta) \right] + \varepsilon Z(\theta) \xi(t)
\label{colnoisephs}
\end{align}
is derived for limit-cycle oscillators driven by colored noise with small correlation time, which depends on the ratio of $\tau_{\xi} / \tau_{\rho}$ and includes both of the limits. Therefore, if we first take $\tau_{\rho} \to 0$ while keeping $\tau_{\xi}$ finite, we obtain the conventional phase Eq.~(\ref{phssde}), while if we first take $\tau_{\xi} \to 0$ and keep $\tau_{\rho}$ finite, we obtain Eq.~(\ref{yoshimu}).

Thus, the frequency of the oscillator may vary when driven by noise. Although the additional term in Eq.~(\ref{colnoisephs}) is formally $O(\varepsilon^2)$, it may give the leading order correction to the phase equation and therefore may not be neglected. For example, if we consider the effect of an additional smooth forcing $f(t)$, we should assume its intensity to be $O(\varepsilon^2)$ rather than $O(\varepsilon)$ in order for $f(t)$ to have comparable effects as $\xi(t)$, because $\xi(t)$ is Gaussian-white. Thus, the additional term may affect synchronization dynamics of noisy oscillators.
On the other hand, it can also be shown that, by the application of the near-identity transform, Eq.~(\ref{colnoisephs}) can be transformed back to the form of the conventional phase equation~(\ref{phssde}), with a slightly different frequency $\tilde{\omega}$. Therefore, once we ``renormalize'' the effect of the noise into the frequency, the conventional phase equation remains still valid.
The analysis in Ref.~\cite{Teramae3} was further developed in Ref.~\cite{Goldobin2}, where an effective white-noise phase equation for limit-cycle oscillators driven by general colored noise was derived.

\subsection{Phase reduction of rhythmic patterns in spatially extended systems}

We have so far considered limit-cycle oscillators described by ODEs.
However, there are also many kinds of rhythmic phenomena described by partial differential equations, such as oscillatory chemical patterns, which are modeled by reaction-diffusion equations, and oscillatory convection, which are described by fluid equations.
In our recent works~\cite{Nakao2,Kawamura}, we developed a phase reduction theory for such stable rhythmic spatiotemporal patterns, namely, for limit cycles in spatially extended systems.
Here, we briefly review the results for reaction-diffusion systems.
Some examples of rhythmic patterns in reaction-diffusion systems are the localized oscillating spots in spatially 1D systems and the target and spiral waves in spatially 2D systems~\cite{Nakao2}.

Consider a reaction-diffusion system described by
\begin{align}
	\frac{\partial}{\partial t} {\bf X}({\bf r}, t) = {\bf F}({\bf X}, {\bf r}) + {\rm D} \nabla^{2} {\bf X},
	\label{rd}
\end{align}
where ${\bf X}({\bf r}, t)$ is a vector field representing the concentrations of chemical species at point ${\bf r}$ and at time $t$,
${\bf F}({\bf X}, {\bf r})$ represents the reaction kinetics of ${\bf X}$ at ${\bf r}$, $\nabla^{2} {\bf X}$ represents the diffusion, and ${\rm D}$ is the matrix of diffusion constants.
We assume that Eq.~(\ref{rd}) exhibits a stable limit-cycle solution, ${\bf X}_{0}({\bf r}, t)$, of temporal period $T$, which satisfies ${\bf X}_{0}({\bf r}, t+T) = {\bf X}_{0}({\bf r}, t)$.
We then introduce a phase $0 \leq \theta < 2\pi$ along the limit-cycle solution, and denote the state of the system as ${\bf X}_{0}({\bf r} ; \theta)$.
The state space of this system is infinite-dimensional, because the 
dynamical variable is the spatially extended vector field ${\bf X}({\bf r}, t)$.
Therefore, the conventional theory for ODEs cannot be applied directly.

In Ref.~\cite{Nakao2}, a phase reduction theory for the reaction-diffusion systems exhibiting rhythmic pattern dynamics is developed.
It can be shown that the phase sensitivity function of the rhythmic pattern ${\bf X}_{0}({\bf r} ; \theta)$ is given by a vector field ${\bf Z}({\bf r}; \theta)$, which is a $2\pi$-periodic function of the phase $\theta$ of the rhythmic pattern.
This space-dependent phase sensitivity function ${\bf Z}({\bf r} ; \theta)$ can be calculated as a $2\pi$-periodic solution to the adjoint partial differential equation
\begin{align}
	\omega \frac{\partial}{\partial \theta} {\bf Z}({\bf r}; \theta)
	= - {\rm J}({\bf r}; \theta)^{\dag} {\bf Z}({\bf r}; \theta) - {\rm D}^{\dag} \nabla^{2} {\bf Z}({\bf r}; \theta),
\label{adjointRD}
\end{align}
where ${\rm J}({\bf r}; \theta)$ is the Jacobi matrix of ${\bf F}({\bf X})$ estimated at ${\bf X} = {\bf X}_{0}({\bf r};\theta)$.
The normalization condition is given by
\begin{align}
	\left\langle {\bf Z}({\bf r};\theta), \frac{\partial}{\partial \theta} {\bf X}_{0}({\bf r};\theta) \right\rangle = 1,
\label{normalizeRD}
\end{align}
where the inner product between two vector fields, ${\bf A}({\bf r})$ and ${\bf B}({\bf r})$, is defined as
$\left\langle {\bf A}({\bf r}), {\bf B}({\bf r}) \right\rangle = \int {\bf A}({\bf r}) \cdot {\bf B}({\bf r})  d{\bf r}$.
We can see that the adjoint equation~(\ref{adjointRD}) and the normalization condition~(\ref{normalizeRD}) are direct generalizations of the corresponding equations~(\ref{adjoint}) and~(\ref{normalization}) for ODEs.

Once the phase sensitivity function ${\bf Z}({\bf r};\theta)$ is obtained, the dynamics of the rhythmic pattern of the reaction-diffusion system subjected to weak perturbations, described by
\begin{align}
	\frac{\partial}{\partial t} {\bf X}({\bf r}, t) = {\bf F}({\bf X}, {\bf r}) + {\rm D} \nabla^{2} {\bf X} + \varepsilon {\bf p}({\bf r}, t),
\end{align}
can be reduced to a one-dimensional phase equation
\begin{align}
	\frac{d}{dt} \theta(t) = \omega + \left\langle {\bf Z}({\bf r};\theta), {\bf p}({\bf r}, t) \right\rangle.
\end{align}
Using this phase equation, we can analyze the synchronization properties of the rhythmic pattern in exactly the same way as for the ordinary limit-cycle oscillators.

\begin{figure}[bt]
	\centering
	\includegraphics[width=\hsize,clip]{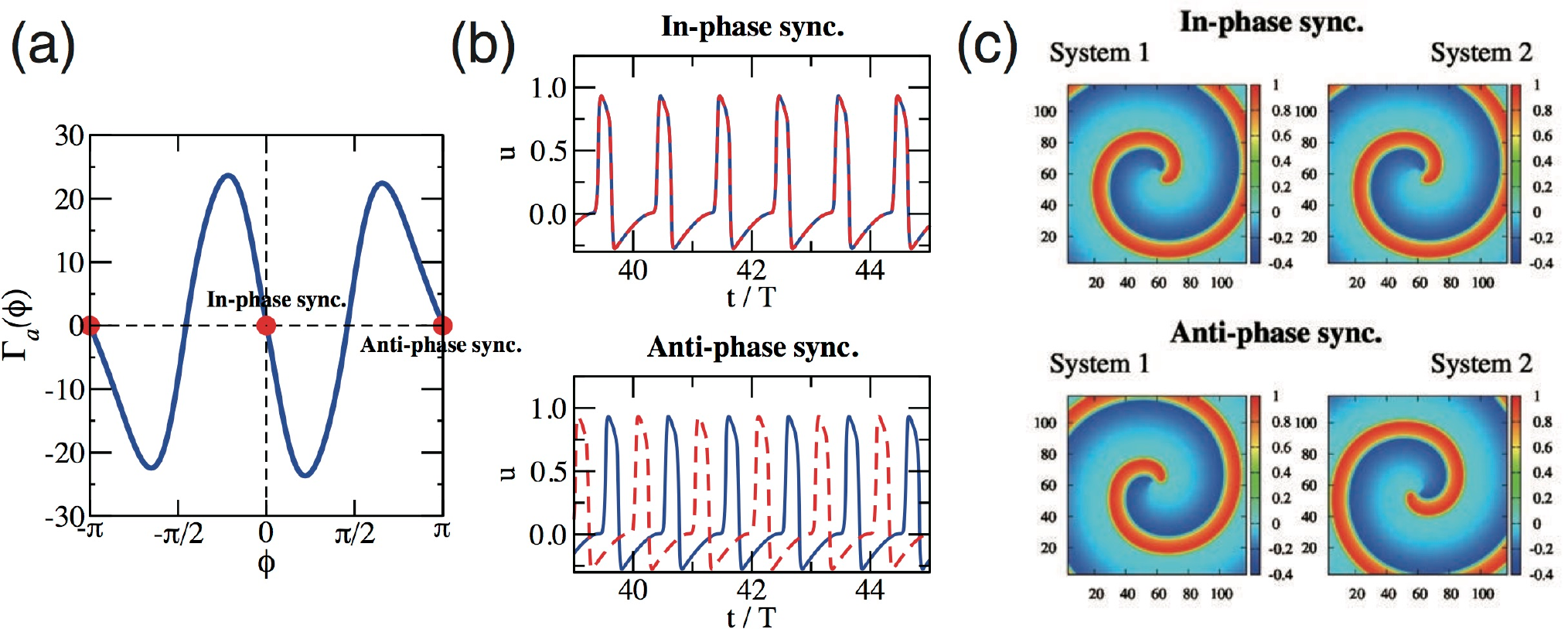}
	\caption{In-phase and anti-phase synchronization of two coupled spiral waves in the FitzHugh-Nagumo reaction-diffusion system. (a) Antisymmetric part of the phase coupling function $\Gamma_{a}(\phi)$. (b) Evolution of the activator variables of the two systems in the in-phase and anti-phase synchronized states. (c) Snapshots of the in-phase and anti-phase synchronized spirals. (From~\cite{Nakao2}).}
	\label{rdsync}
\end{figure}

As an example, mutual synchronization of a pair of periodically rotating spiral waves is shown in Fig.~\ref{rdsync}. These spirals are described by symmetrically coupled FitzHugh-Nagumo reaction-diffusion systems, 
\begin{align}
	\frac{\partial}{\partial t} {\bf X}_1({\bf r}, t) &= {\bf F}({\bf X}_1, {\bf r}) + {\rm D} \nabla^{2} {\bf X}_1 + \varepsilon ( {\bf X}_2 - {\bf X}_1 ), \cr
	\frac{\partial}{\partial t} {\bf X}_2({\bf r}, t) &= {\bf F}({\bf X}_2, {\bf r}) + {\rm D} \nabla^{2} {\bf X}_2 + \varepsilon ( {\bf X}_1 - {\bf X}_2 ),
\label{cplrds}
\end{align}
where ${\bf X}_1$ and ${\bf X}_2$ represent the states of the system 1 and 2, respectively, and the last terms represent weak inter-system diffusive coupling with intensity $\varepsilon$.
By applying the phase reduction to Eq.~(\ref{cplrds}), a pair of coupled phase equations~(\ref{twocplphs}) is obtained, where the phase coupling function $\Gamma(\phi)$ can be calculated from the limit-cycle solution ${\bf X}_{0}({\bf r};\theta)$ and phase sensitivity function ${\bf Z}({\bf r};\theta)$ corresponding to the spiral waves.

The dynamics of the phase difference between the two spirals is then determined by Eq.~(\ref{2oscphsdif}). From the antisymmetric component of $\Gamma_{a}(\phi)$ in Fig.~\ref{rdsync}(a), it is expected that both in-phase and anti-phase synchronized states are stable.
Indeed, the numerical simulations show either in-phase or anti-phase synchronized spirals depending on the initial phase difference.
Figure~\ref{rdsync}(b) plots temporal evolution of the activator variables of the two reaction-diffusion systems measured at the corresponding locations, showing in-phase and anti-phase synchronization.
Figure~\ref{rdsync}(c) shows the snapshots of in-phase and anti-phase synchronized spirals.
See~\cite{Nakao2,Kawamura} for the details of the theory, simulation, and other examples of rhythmic spatiotemporal patterns.

\subsection{Other recent topics in phase reduction theory}

As we explained above for the reaction-diffusion systems, it is possible, {\em in principle}, to develop phase reduction theory for dynamical systems as long as they possess stable limit-cycle solutions. Though there are problem-specific technical difficulties, studies in this direction are in progress recently.
Also, inverse methods for inferring the phase models from experimental data without the knowledge of the detailed mathematical models have been developed recently. Here we mention some of such advances in phase reduction theory.

(i) {\em Phase reduction of collective oscillations in coupled oscillators.}
When a population of nonlinear oscillators undergoes mutual synchronization and exhibits stable collective oscillations, they can be considered as a single macroscopic oscillator~\cite{Kawamura2,Kori}.
In Ref.~\cite{Kawamura2}, the collective phase sensitivity function of the macroscopic oscillation is derived for a population of noisy limit-cycle oscillators with global coupling.
By analyzing the time-periodic solution of the nonlinear Fokker-Planck equation, macroscopic phase sensitivity of the collective oscillations is derived from the microscopic phase sensitivity of the individual oscillators.
Also, in Ref.~\cite{Kori}, the collective phase sensitivity function is obtained for a population of nonidentical phase oscillators coupled via a general network, which exhibits fully phase-locked collective oscillations. It was shown that the collective phase sensitivity can largely differ depending on the network structure.
As in the case of the reaction-diffusion system, once we know the phase sensitivity function, we can analyze the collective oscillations of a population of coupled oscillators by a single phase equation.

(ii) {\em Phase reduction of delay-induced oscillations.}
There are various examples of real-world systems with delays, e.g., in nonlinear optics and biological regulations, described by a delay-differential equation of the form $dx(t)/dt = F(x(t), x(t-\tau))$, where $\tau$ is the delay time. The delay often induces limit-cycle oscillations~\cite{Glass}. Also, limit-cycle orbits stabilized by delayed-feedback control of chaotic systems are described by delay-differential equations~\cite{Novicenko}.
Such delay-differential equations are infinite-dimensional dynamical systems, because the whole information of the variable $x$ in the interval $[t-\tau, t]$ is required to specify the system state.
Recently, Novi\v{c}enko and Pyragas~\cite{Novicenko} and also Kotani {\it et al.}~\cite{Kotani} developed phase reduction theory for delay-induced limit-cycle oscillations.
By introducing appropriate ``bilinear form'' (inner product) defining the distance between the infinite-dimensional oscillator states, the adjoint equation for the phase sensitivity function of delay-induced oscillations is derived.
Reflecting the complex waveform due to time delay, interesting synchronization dynamics, such as multi-modal phase locking~\cite{Kotani}, can be observed in such systems.

(iii) {\em Phase reduction of hybrid limit-cycle oscillators.}
In modeling real-world phenomena, piecewise-smooth dynamical systems with sudden jumps or non-smoothness of the state variables are widely used, for example, in the models of electric circuits with discontinuous switching or animal gaits with sudden collisions with the ground. 
Such dynamical systems are often called hybrid dynamical systems, and can also exhibit stable limit-cycle oscillations.
One of the simplest example of the piecewise-smooth oscillator is the integrate-fire-type model of neurons, in which the membrane potential of the neuron monotonically increases until it reaches a threshold value, and then it is discontinuously reset to the resting value.
The phase reduction has also been applied also to such models~\cite{Brown}, and general phase reduction theory for hybrid limit cycles can also be developed~\cite{Shirasaka1}.
Due to the non-smoothness, hybrid oscillators can also exhibit nonconventional, interesting synchronization properties.

(iv) {\em Non-autonomous oscillators and strongly perturbed oscillators.}
In this article, we have considered only autonomous oscillators. However, in the real world, the oscillator may also be affected by time-varying environmental factors and, for example, the frequency of the oscillation may vary with time. To analyze such non-autonomous oscillators, Stefanovska and coworkers recently introduced the notion of the ``chronotaxic system''~\cite{Chronotaxic1,Chronotaxic2,Chronotaxic3}, described by ODEs of the form $\dot{\bf p} = {\bf f}({\bf p})$
and $\dot{\bf x} = {\bf g}({\bf x}, {\bf p})$, where ${\bf p}$ represents the environmental
factors and ${\bf x}$ the oscillator state.
They showed that the heart rate variability measured from human subjects can be successfully
interpreted as a chronotaxic system rather than as an ordinary autonomous limit-cycle oscillator,
and also studied synchronization of coupled chronotaxic oscillators~\cite{Chronotaxic2}.

In a recent independent study~\cite{Strong}, Kurebayashi {\em et al.} proposed a generalization of phase reduction theory to strongly perturbed oscillators (see also Park and Ermentrout~\cite{Park} for an extension). By decomposing the perturbation into a slowly varying component and remaining weak fluctuations, and by considering a continuous family of limit-cycle orbits parametrized by the slow component, a generalized phase equation valid for oscillators subjected to largely varying perturbations is derived, and has successfully been applied to the analysis of oscillators driven by strong periodic forcing.
By identifying the slow component of the perturbation as an environmental factor ${\bf p}$, this generalized phase reduction theory could be utilized to analyze synchronization properties of non-autonomous chronotaxic oscillators.

(v) {\em Inference of phase models from experimental data.}
Recently, the importance of data-driven approach in dynamical systems has been emphasized\footnote{The main subject of these papers is the Koopman operator, which has an interesting relation to the isochrons of the oscillator. It can also be used in defining amplitude variables of the oscillator in the phase-amplitude description.}~\cite{Mezic,Mauroy}.
Various inverse methods to obtain the phase models from experimental data, which do not rely on the knowledge of detailed mathematical models, have been proposed
~\cite{Kralemann}-\cite{Kralemann2},\cite{Tokuda}-\cite{Ota3}.
For example, in Ref.~\cite{Tokuda}, the phase coupling function has been inferred from time series of electrochemical oscillators, and in Ref.~\cite{Kralemann2}, the phase coupling function as well as the phase sensitivity function of cardiorespiratory interaction have been directly reconstructed from observed data.
In Refs.~\cite{Stankovski,Ota3}, general Bayesian frameworks for estimating the phase coupling functions directly from multivariate experimental data are proposed.
As the complexity of the rhythmic phenomena that we are interested increases, detailed mathematical models will be more difficult to attain.
Such inverse approaches to coupled oscillators will be increasingly important in the future and find many real-world applications.

(vi) {\em Phase-amplitude description of limit-cycle oscillators.}
Though the phase reduction theory has played a dominant role in the analysis of weakly perturbed limit-cycle oscillators, in realistic problems, the oscillators may be subjected to stronger perturbations.
In such cases, effect of the amplitude degrees of freedom of the oscillator, i.e., deviation of the oscillator state from the unperturbed limit cycle, cannot be ignored and the reduced phase equation may yield inaccurate predictions or even break down.
When the perturbation is not too strong, the real part of the second Floquet exponent characterizes how large the effect of the amplitude degrees of freedom is; smaller value indicates faster convergence of the oscillator state to the limit cycle and therefore faster decay of the deviation.
Decomposition of the dynamics around a periodic orbit into phase and amplitude components had been discussed e.g., in a classical textbook by Hale~\cite{Hale}, and such kind of decomposition has recently been used for phase-amplitude description of the dynamics of a perturbed oscillator around the limit cycle~\cite{Wedgwood,Wilson,Shirasaka2}.
Unlike phase reduction, the reduced system is still two-dimensional (or more), but intriguing dynamics that cannot be described by phase-only equations can be analyzed by using phase-amplitude equations.

\section{Summary}

We have briefly introduced phase reduction theory for nonlinear oscillators.
For detailed explanations and derivations of the equations, interested readers are encouraged to consider more advanced literature.
In this introductory article, we wish to emphasize that, through phase reduction theory, universal results that do not depend on the physical or chemical characteristics of the oscillators can be extracted from the original model.
For example, once a pair of oscillators is reduced to coupled phase equations, it does not matter if the oscillators are in fact pendulum clocks or spiking neurons; provided we are interested in their synchronization properties, only their frequencies and phase response properties are relevant.

Indeed, phase reduction has a conceptual importance that is beyond the simplification of the dynamical equations. In many real-world oscillatory phenomena, a complete understanding of the detailed mechanisms of the oscillations is difficult to achieve. For example, experimentally specifying all of the genes, proteins, and interactions that are related to certain biochemical oscillations in a living cell can be a quite challenging task.
However, from the standpoint of phase reduction theory, we can obtain a general 
phenomenological phase model of oscillatory systems by measuring their phase response properties and make quantitative predictions of their dynamics, without knowing their physical or material details.
Moreover, phase reduction often gives similar phase models for physically distinct phenomena, thereby revealing the underlying {\em universal mathematical mechanisms} of seemingly very different systems.
In this sense, phase reduction theory gives an explicit prescription to realize the concept of ``inter-phenomenological understanding of the diverse world on the basis of predicative invariance'' put forward by Kuramoto for various rhythmic systems.
As increasingly large volumes of data are being acquired in complex real-world systems, data-driven methods for dynamical systems, combined with the simplicity and generality of phase reduction theory, will play important roles in the analysis of these systems.
The phase reduction theory will continue to serve as a simple and powerful framework for understanding various rhythmic phenomena in a wide range of science and engineering.

\section*{Acknowledgements}

The author is grateful to Prof. Stefanovska for the invitation to write this introductory article, to Prof. Kuramoto for all of his advice and support, and to Dr. Kawamura, Dr. Arai, Dr. Kurebayashi, Dr. Shirasaka, Prof. Kotani, and Prof. Ermentrout for various discussions.
The author thanks financial support from JSPS (KAKENHI 26120513, 26103510, 25540108) and CREST Kokubu project (JST, 2009-2014).

\appendix

\section*{Appendix}

\section{Linear stability of the limit-cycle solution}
\label{App:Floquet}

Linear stability of the limit-cycle solution ${\bf X}_{0}(t)$ is characterized by its Floquet exponents.
Suppose a slightly perturbed solution ${\bf X}(t) = {\bf X}_{0}(t) + {\bf y}(t)$ from the limit-cycle orbit ${\bf X}_{0}(t)$, where ${\bf y}(t)$ is a small deviation, and plug into Eq.~(\ref{ode}).
The linearized variational equation for ${\bf y}(t)$ is given by
\begin{align}
	\frac{d}{dt} {\bf y}(t) = {\rm J}(t) {\bf y}(t),
	\label{floq1}
\end{align}
where ${\rm J}(t)$ is the Jacobi matrix of ${\bf F}({\bf X})$ at ${\bf X} = {\bf X}_{0}(t)$ whose $(i,j)$ element is ${\rm J}_{ij}(t) = \left. \partial F_{i}({\bf X}) / \partial X_{j} \right|_{{\bf X} = {\bf X}_{0}(t)}$.
Because ${\bf X}_{0}(t)$ is a $T$-periodic solution, ${\rm J}(t)$ is a $T$-periodic matrix.
Therefore, by the Floquet theorem~\cite{GH}, solutions to Eq.~(\ref{floq1}) can be expressed in the form
\begin{align}
	{\bf y}(t) = {\rm P}(t) \exp( t {\rm R} ) {\bf y}(0),
\end{align}
where the matrix ${\rm R}$ is constant and ${\rm P}$ satisfies ${\rm P}(t+T) = {\rm P}(t)$ and ${\rm P}(0) = {\rm I}$. In particular, ${\bf y}(T) = \exp( T {\rm R} ) {\bf y}(0)$ holds.
We introduce the eigenvalues and associated eigenvectors of the matrix $\exp( T {\rm  R})$ as $\lambda_{i} = \exp( {\mu_{i}} )$ and ${\bf e}_{i}$ ($i=1, ..., M$), i.e., $\exp( T {\rm  R}) {\bf e}_{i} = \lambda_{i} {\bf e}_{i}$, where $\lambda_{i}$ and $\mu_{i}$ are called the Floquet multipliers and exponents, respectively.
By decomposing the initial deviation as
\begin{align}
	{\bf y}(0) = \sum_{i=1}^{M} a_{i} {\bf e}_{i}
\end{align}
where $a_{1}, ..., a_{M}$ are expansion coefficients, the deviation ${\bf y}(T)$ after one period of oscillation is given by
\begin{align}
{\bf y}(T) = \exp( T {\rm R} ) {\bf y}(0) = \sum_{i=1}^{M} a_{i} \exp( T {\rm R} ) {\bf e}_{i}
= \sum_{i=1}^{M} \lambda_{i} a_{i} {\bf e}_{i} = \sum_{i=1}^{M} e^{\mu_{i}} a_{i} {\bf e}_{i}.
\end{align}
Because infinitesimal perturbation along the limit cycle is neutrally stable and remains constant after one period of oscillation, one of the Floquet exponents vanishes, i.e., $\mu_{1} = 0$ ($\lambda_{1} = 1$).
The limit cycle is linearly stable if the real parts of all other Floquet exponents are negative, i.e., $\mbox{Re}~\mu_{2}, ..., \mbox{Re}~\mu_{M} < 0$ ($\lambda_{2}$, ..., $\lambda_{M}$ lie inside the unit circle on the complex plane), because then the deviation from the limit-cycle orbit eventually decays.

\section{The Stuart-Landau oscillator}
\label{App:SL}

It is generally difficult to obtain explicit analytical expressions for the limit-cycle orbit and related quantities.
For the {\em Stuart-Landau} (SL) oscillator, which is the normal form of the supercritical Hopf bifurcation~\cite{Winfree,Kuramoto,Ermentrout1,Strogatz0,Brown}, the limit-cycle orbit, phase function, and phase sensitivity functions can be analytically calculated as follows.

The SL oscillator has a two-dimensional state variable ${\bf X} = (X, Y)$, which obeys
\begin{align}
\frac{d}{dt} \left( \begin{array}{c} X(t) \\ Y(t) \end{array} \right)
=
\left( \begin{array}{c}
X - \alpha Y - ( X - \beta Y ) ( X^{2} + Y^{2} )
\\
\alpha X + Y - ( \beta X + Y ) ( X^{2} + Y^{2} )
\end{array} \right),
\end{align}
where $\alpha$ and $\beta$ are real parameters.
By introducing a complex variable $W = X + i Y$, the above equation can also be expressed as
\begin{align}
	\frac{d}{dt} W(t) = ( 1 + i \alpha ) W - ( 1 + i \beta ) |W|^{2} W.
\end{align}
Moreover, by introducing modulus $R = |W| = \sqrt{X^{2}+Y^{2}}$ and argument $\phi = \arctan Y / X$ of $W = X + i Y$ satisfying $W = R \exp({i \phi})$, $R$ and $\phi$ obey
\begin{align}
	\frac{d}{dt} R(t) = R - R^{3}, \quad \frac{d}{dt} \phi(t) = \alpha - \beta R^{2}.
	\label{SLpolar}
\end{align}
Note that this $\phi$ is merely the ``angle'' of $W$ and not the ``phase'' in the present context of phase reduction theory, and is defined only when $R>0$.

From the above equations, it is obvious that $R(t) \to 1$ as long as $R(0) > 0$ initially, so the limit-cycle orbit $\chi$ can be expressed as (assuming $W(0) = 1$, i.e., $R(0)=1$ and $\phi(0)=0$ without loss of generality)
\begin{align}
R(t) = 1, \quad \phi(t) = (\alpha - \beta) t,
\end{align}
namely, 
\begin{align}
	W_{0}(t) = \exp [ i( \alpha - \beta ) t ]
\end{align}
or
\begin{align}
	{\bf X}_{0}(t) = (X_{0}(t), Y_{0}(t)) =	(\cos ( \alpha - \beta ) t,\ \sin ( \alpha - \beta ) t),
\end{align}
which represents a unit circle. The frequency and period of the oscillation are given by $\omega = \alpha - \beta$ and $T = 2 \pi / (\alpha - \beta)$, respectively.

The phase function of the above SL oscillator is given by
\begin{align}
	\Theta(W) = \Theta(R, \phi) = \mbox{arg} W - \beta \ln |W| = \phi - \beta \ln R.
\end{align}
Indeed, by using Eq.~(\ref{SLpolar}), the phase $\theta(t) = \Theta(W(t))$ of the SL oscillator obeys
\begin{align}
\frac{d}{dt} \theta(t) = \frac{d}{dt} \phi(t) - \beta \frac{1}{R(t)} \frac{d}{dt} R(t) = \alpha - \beta R(t)^{2} - \beta ( 1 - R(t)^{2} ) = \alpha - \beta = \omega,
\end{align}
so that $\theta(t)$ increases with a constant frequency $\omega$ not only on the limit-cycle orbit $\chi$ but in its basin of attraction.
The limit-cycle orbit $\chi$ can be expressed as
\begin{align}
W_{0}(\theta) = \exp ( i \theta )
\end{align}
or
\begin{align}
{\bf X}_{0}(\theta) = (X_{0}(\theta), Y_{0}(\theta)) = ( \cos \theta, \sin \theta)
\label{sllc}
\end{align}
as a function of the phase variable $\theta$ ($0 \leq \theta < 2\pi$).

Using the phase function, the phase response function to an impulsive stimulus $z = x + i y$ in the complex plane, which kicks the oscillator state from $W_{0}(\theta)$ to $W = W_{0}(\theta) + z$ (or from $(X_0(\theta), Y_0(\theta))$ to $(X_0(\theta)+x, Y_0(\theta)+y)$ in the $(X,Y)$ plane), can be expressed as
\begin{align}
	g(\theta ; z=x+iy) = \Theta(W_{0}(\theta) + z) - \Theta(W_{0}(\theta)) = \Theta(W_{0}(\theta) + x + i y) - \theta.
\end{align}
Moreover, the phase sensitivity function ${\bf Z}(\theta) = (Z_{X}(\theta), Z_{Y}(\theta))$ can be obtained by differentiating $\Theta(W)$ by $x$ and $y$ as
\begin{align}
{\bf Z}(\theta) &= (Z_{X}(\theta), Z_{Y}(\theta)) 
\cr
&= \mbox{grad}_{(X,Y) = (X_{0}(\theta), Y_{0}(\theta))} \Theta(W)
= \left( \frac{\partial g(\theta ; z)}{\partial x}, \frac{\partial g(\theta ; z)}{\partial y} \right)_{(x, y) = (0, 0)},
\end{align}
which yields
\begin{align}
{\bf Z}(\theta) = (- \sin \theta - \beta \cos \theta, \quad \cos \theta - \beta \sin \theta).
\label{slZ}
\end{align}
Thus, the SL oscillator has a circular limit-cycle orbit and a sinusoidal phase sensitivity function.

\begin{figure}[bt]
	\centering
	\includegraphics[width=\hsize,clip]{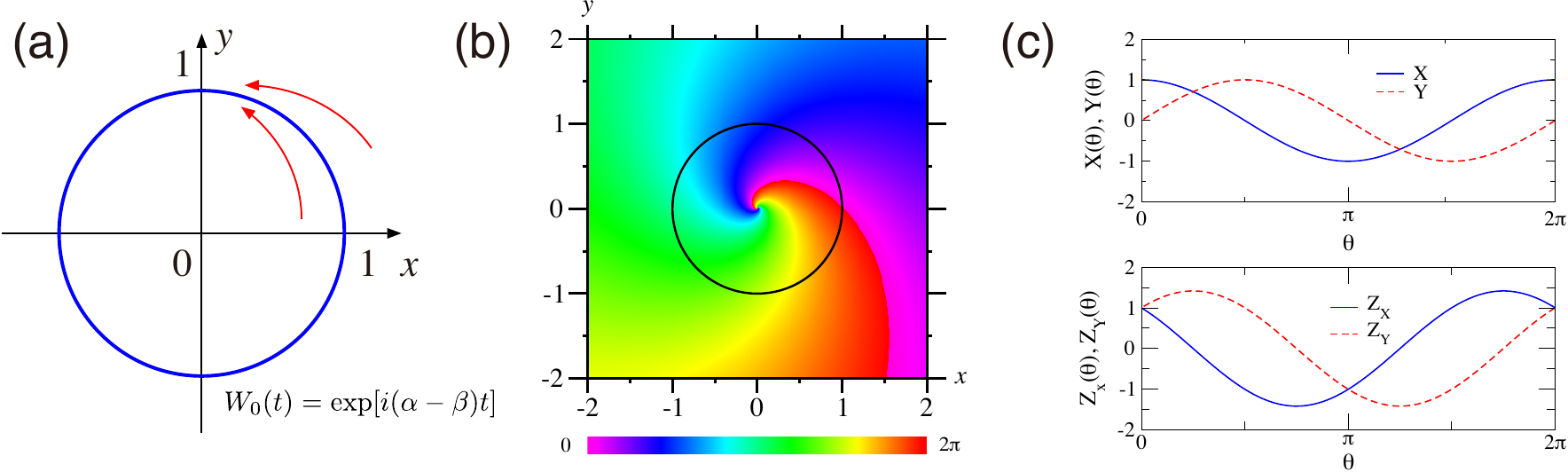}
	\caption{(a) Limit-cycle orbit of the Stuart-Landau oscillator. (b) Phase function $ \Theta(X,Y)$. (c) Limit-cycle orbit $(X_{0}(\theta), Y_{0}(\theta))$ and the corresponding phase sensitivity function $(Z_{X}(\theta), Z_{Y}(\theta))$. The parameter values are $\alpha=1$ and $\beta=-1$.}
	\label{figA1}
\end{figure}

Let us now consider a pair of SL oscillators with weak mutual coupling, described by
\begin{align}
	\frac{d}{dt} W_{1}(t) &= ( 1 + i \alpha ) W_{1} - ( 1 + i \beta ) |W_{1}|^{2} W_{1} + \varepsilon ( 1 + i \delta ) ( W_{2} - W_{1} ), \cr
	\frac{d}{dt} W_{2}(t) &= ( 1 + i \alpha ) W_{2} - ( 1 + i \beta ) |W_{2}|^{2} W_{2} + \varepsilon ( 1 + i \delta ) ( W_{1} - W_{2} ),
\end{align}
where $W_{1}$ and $W_{2}$ are the complex variables of the oscillators $1$ and $2$, respectively, $\varepsilon > 0$ is a small parameter representing the coupling intensity, and $\delta$ is a real coupling parameter.
The above equations generally describe a pair of coupled oscillators near the supercritical Hopf bifurcation~\cite{Kuramoto}.
Note that the coupling function is given in the real expression as
\begin{align}
{\bf G}({\bf X}_{1}, {\bf X}_{2}) = \left( \begin{array}{cc} 1 & -\delta \cr \delta & 1 \end{array} \right) \left( \begin{array}{c} X_{2} - X_{1} \cr Y_{2} - Y_{1} \end{array} \right).
\end{align}
By applying the phase reduction and averaging, the approximate phase equations are given by Eq.~(\ref{twocplphs}), where the phase coupling function can be calculated as
\begin{align}
\Gamma(\varphi) &=  \frac{1}{2\pi} \int_{0}^{2\pi} {\bf Z}(\varphi + \psi) \cdot 
{\bf G}( {\bf X}_{0}(\varphi + \psi), {\bf X}_{0}(\psi) ) d\psi
\cr
&=
(\beta - \delta) ( 1 - \cos \varphi) - (1 + \beta \delta) \sin \varphi
\end{align}
using the limit-cycle solution~(\ref{sllc}) and phase sensitivity function~(\ref{slZ}).
The antisymmetric part of this phase coupling function is given by
\begin{align}
\Gamma_{a}(\varphi) = \Gamma(\varphi) - \Gamma(-\varphi) = - 2 ( 1 + \beta \delta) \sin \varphi,
\end{align}
and, therefore, the in-phase synchronized state $\varphi = 0$ is stable and the anti-phase synchronized state $\varphi = \pm \pi$ is unstable when $1 + \beta \delta > 0$, and vice versa when $1 + \beta \delta < 0$.
This condition generally arises in the analysis of the uniformly synchronized state of a network of diffusively coupled Stuart-Landau oscillators, and is called the Benjamin-Feir condition~\cite{Kuramoto,Nakao3,Nakagawa,Aranson,NakaoCGL}. Instability of the uniformly synchronized state often leads to nontrivial dynamics of the oscillators including chaos.

\begin{figure}[tb]
	\centering
	\includegraphics[width=\hsize,clip]{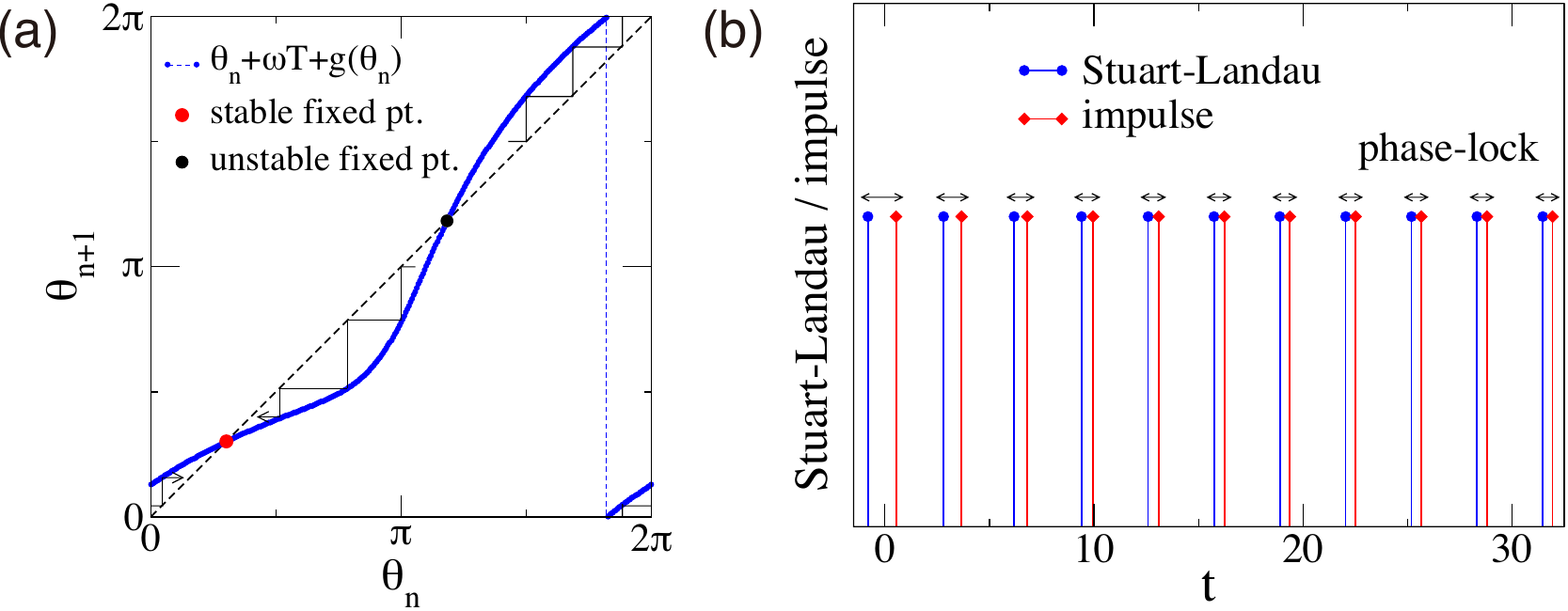}
	\caption{(a) Phase map of the Stuart-Landau oscillator with $\alpha=1$, $\beta=-1$, where the impulse of amplitude $0.5$ is applied in the $x$ direction. (b) Phase locking of the Stuart-Landau oscillator to periodic impulses of period $T = \pi$. The zero-crossing events of the oscillators (from $Y<0$ to $Y>0$) and the impulses are shown by vertical lines. The phase difference converges to a fixed value given by the stable fixed point of the phase map. }
	\label{fig-phasemap}
\end{figure}

\section{Phase equation for impulse-driven oscillators}
\label{App:Impulse}

Suppose a limit-cycle oscillator subjected to external impulses described by
\begin{align}
\frac{d}{dt} {\bf X}(t) = {\bf F}({\bf X}) + \sum_{n} {\bf I}_{n} \delta(t - t_{n}),
\end{align}
where ${\bf I}_{n}$ represents the direction and intensity of the $n$th impulse and $t_{n}$ its arrival time.
It is not necessary that the impulses ${\bf I}_{n}$ are weak, but the intervals between the impulses should be sufficiently long
so that the oscillator state perturbed by an impulse can return sufficiently close to the limit-cycle orbit before receiving the next impulse.
Then, the above equation can be reduced to a phase equation of the form
\begin{align}
\frac{d}{dt} \theta(t) = \omega + \sum_{n} g(\theta(t_{n}-0) ; {\bf I}_{n}) \delta(t - t_{n}).
\end{align}
Moreover, by denoting the phase $\theta(t)$ of the oscillator just before the $n$th impulse as $\theta_{n} = \theta(t_{n} - 0)$, the above equation can be simplified to the following {\em phase map}~\cite{Glass}:
\begin{align}
\theta_{n+1} = \theta_{n} + \omega (t_{n+1} - t_{n}) + g(\theta_{n}).
\label{phasemap}
\end{align}
When the impulse is periodically applied with period (i.e., interval) $T_{ext}$, this equation reduces to
\begin{align}
\theta_{n+1} = \theta_{n} + \omega T + g(\theta_{n}).
\end{align}
Thus, the dynamics of the limit-cycle oscillator driven by periodic impulses can be predicted by analyzing the above simple one-dimensional map.
As shown in Fig.~\ref{fig-phasemap}, if this phase map has a stable fixed point, the oscillator is phase-locked to the external impulses.
As discussed in Ref.~\cite{Arai}, synchronization of limit-cycle oscillators driven by common random impulses can also be analyzed by using the phase map~(\ref{phasemap}), with $t_{n+1} - t_{n}$ given by a random variable representing the intervals between the random impulses.

\section{Derivation of the adjoint equation}
\label{App:adjoint}

We here derive the adjoint equation, based on the simple argument by Brown, Moehlis, and Holmes~\cite{Brown}. The following derivation is from~\cite{Nakao2}, which uses the same idea to derive the adjoint equation for reaction-diffusion systems.
More mathematical proof is based on the Fredholm alternative theory, which gives the phase equation as the solvability condition of the linearized variational equation of periodically perturbed oscillators, where ${\bm Z}(\theta)$ spans the nullspace of the linear variational operator and satisfies the adjoint equation~\cite{Hoppensteadt,Ermentrout1,Hale}.

Let us consider an oscillator state ${\bf X}$ near the oscillator state ${\bf X}_{0}(\theta)$ on the limit-cycle orbit $\chi$ with phase $\theta$.
When $| {\bf X} - {\bf X}_{0}(\theta) |$ is sufficiently small, the phase $\Theta({\bf X})$ of ${\bf X}$ can be linearly approximated as (see Sec.~3)
\begin{align}
\Theta({\bf X})
&= \Theta({\bf X}_{0}(\theta) + {\bf X} - {\bf X}_{0}(\theta)) \cr
&\simeq \Theta({\bf X}_{0}(\theta)) + \mbox{grad}_{{\bf X} = {\bf X}_{0}(\theta)} \Theta({\bf X}) \cdot ({\bf X} - {\bf X}_{0}(\theta)) \cr
&= \theta + {\bf Z}(\theta) \cdot ( {\bf X} - {\bf X}_{0}(\theta) ),
\label{phstaylor}
\end{align}
where the error is $O(|{\bf X} - {\bf X}_{0}(\theta)|^{2})$.
Suppose that we prepare an unperturbed oscillator state ${\bf X}_{0}(\theta(t))$ with phase $\theta(t)$ on $\chi$ and a slightly perturbed oscillator state ${\bf X}(t) = {\bf X}_{0}(\theta(t)) + {\bf u}(t)$ from $\chi$, and evolve them without further external perturbations, i.e., by $d{\bf X}(t)/dt  = {\bf F}({\bf X})$.
The small variation ${\bf u}(t) = {\bf X}(t) - {\bf X}_{0}(\theta(t))$ obeys a linearized equation
\begin{align}
\frac{d}{dt} {\bf u}(t) = {\rm J}(\theta(t)) {\bf u}(t),
\end{align}
where ${\rm J}(\theta(t))$ is the Jacobi matrix of ${\bf F}({\bf X})$ at ${\bf X} = {\bf X}(\theta(t))$.
During this evolution, the phase difference between the oscillators should remain constant, namely,
\begin{align}
\frac{d}{dt} [ \Theta({\bf X}(t)) - \Theta({\bf X}_{0}(\theta(t))) ]
=
\frac{d}{dt} [ \Theta({\bf X}(t)) - \theta(t) ]
= 0,
\label{constphase}
\end{align}
because no external perturbation is applied.

Plugging ${\bf X}(t) = {\bf X}_{0}(\theta(t)) + {\bf u}(t)$ into $\Theta({\bf X}(t))$ and using the linear approximation of $\Theta({\bf X})$ in Eq.~(\ref{phstaylor}), Eq.~(\ref{constphase}) is transformed to
\begin{align}
	0 = \frac{d}{dt} \left[ {\bf Z}(\theta(t)) \cdot {\bf u}(t) \right] 
	&= \frac{d{\bf Z}(\theta(t))}{dt} \cdot {\bf u}(t) + {\bf Z}(\theta(t)) \cdot \frac{d{\bf u}(t)}{dt} \cr
	&= \frac{d\theta(t)}{dt} \left.\frac{d{\bf Z}(\theta)}{d\theta}\right|_{\theta=\theta(t)} \cdot {\bf u}(t) + {\bf Z}(\theta(t)) \cdot {\rm J}(\theta(t)) {\bf u}(t) \cr
	&= \left[ \omega \left.\frac{d{\bf Z}(\theta)}{d\theta}\right|_{\theta=\theta(t)} + {\rm J}^{\dag}(\theta(t)) {\bf Z}(\theta(t)) \right] \cdot {\bf u}(t).
\end{align}
Because this equation should hold for arbitrary ${\bf u}(t)$, the following relation should be satisfied:
\begin{align}
	\omega \frac{d}{d\theta} {\bf Z}(\theta) = - {\rm J}^{\dag}(\theta) {\bf Z}(\theta)
	\label{adjointderiv}
\end{align}
for $0 \leq \theta < 2\pi$.
If ${\bf Z}(\theta)$ satisfies this equation, the phase $\Theta({\bf X}(t))$ of the initially perturbed state and the phase $\theta(t) = \Theta({\bf X}_{0}(\theta(t)))$ of the unperturbed state remain constant to the first order in ${\bf X}(t) - {\bf X}_{0}(\theta(t))$, satisfying the assumption for the phase function to the linear order.
Equation~(\ref{adjointderiv}) is the adjoint equation for the phase sensitivity function ${\bf Z}(\theta)$.

Because the adjoint equation is linear, appropriate normalization of ${\bf Z}(\theta)$ is necessary.
The normalization condition can be obtained by differentiating the identity $\theta(t) = \Theta({\bf X}_{0}(\theta(t))$ by $t$ as
\begin{align}
	\omega = \frac{d\theta(t)}{dt} = \mbox{grad}_{{\bf X} = {\bf X}_{0}(\theta(t))} \Theta({\bf X}) \cdot \frac{d{\bf X}_{0}(\theta(t))}{dt} = {\bf Z}(\theta(t)) \cdot {\bf F}({\bf X}_{0}(\theta(t)),
\end{align}
namely,
\begin{align}
{\bf Z}(\theta) \cdot {\bf F}({\bf X}_{0}(\theta)) = \omega
\end{align}
should be satisfied for all $0 \leq \theta < 2\pi$. From this equation, the normalization condition Eq.~(\ref{normalization}) is obtained by using $d{\bf X}_{0}(t)/dt = \omega d{\bf X}_{0}(\theta)/d\theta = {\bf F}({\bf X}_{0}(\theta))$.

\section{Numerical algorithm for the phase sensitivity function}
\label{App:Numerics}

To obtain the phase sensitivity function ${\bf Z}(\theta)$ numerically, it is convenient to evolve the adjoint equation~(\ref{adjointderiv}) backward in time.
Since we assume that the limit-cycle orbit $\chi$ is linearly stable, the real parts of the Floquet exponents of $\chi$ are all negative
(except for the zero exponent corresponding to the direction along the orbit).
Therefore, the adjoint equation~(\ref{adjointderiv}) is linearly unstable when evolved forward in time, because of the minus sign before ${\rm J}^{\dag}(\theta)$.
This makes it difficult to obtain the time-periodic solution ${\bf Z}(\theta+2\pi) = {\bf Z}(\theta)$ by numerically evolving Eq.~(\ref{adjointderiv}), because it is a neutrally stable mode and therefore easily buried in other unstable modes that grow quickly.
However, if Eq.~(\ref{adjointderiv}) is evolved {\em backward in time} for sufficiently long time, only the neutrally stable periodic mode ${\bf Z}(\theta+2\pi) = {\bf Z}(\theta)$ remains, because all other modes are now linearly stable and decay quickly.
Thus, by recording the oscillator state ${\bf X}_{0}(\theta)$ on $\chi$ for one period of oscillation, and by evolving the adjoint equation backward in time using the recorded oscillator state for a long time, ${\bf Z}(\theta)$ can be obtained numerically. This is called the ``adjoint method'' after Ermentrout~\cite{Ermentrout2}.

\begin{figure}[tb]
	\centering
	\includegraphics[width=\hsize,clip]{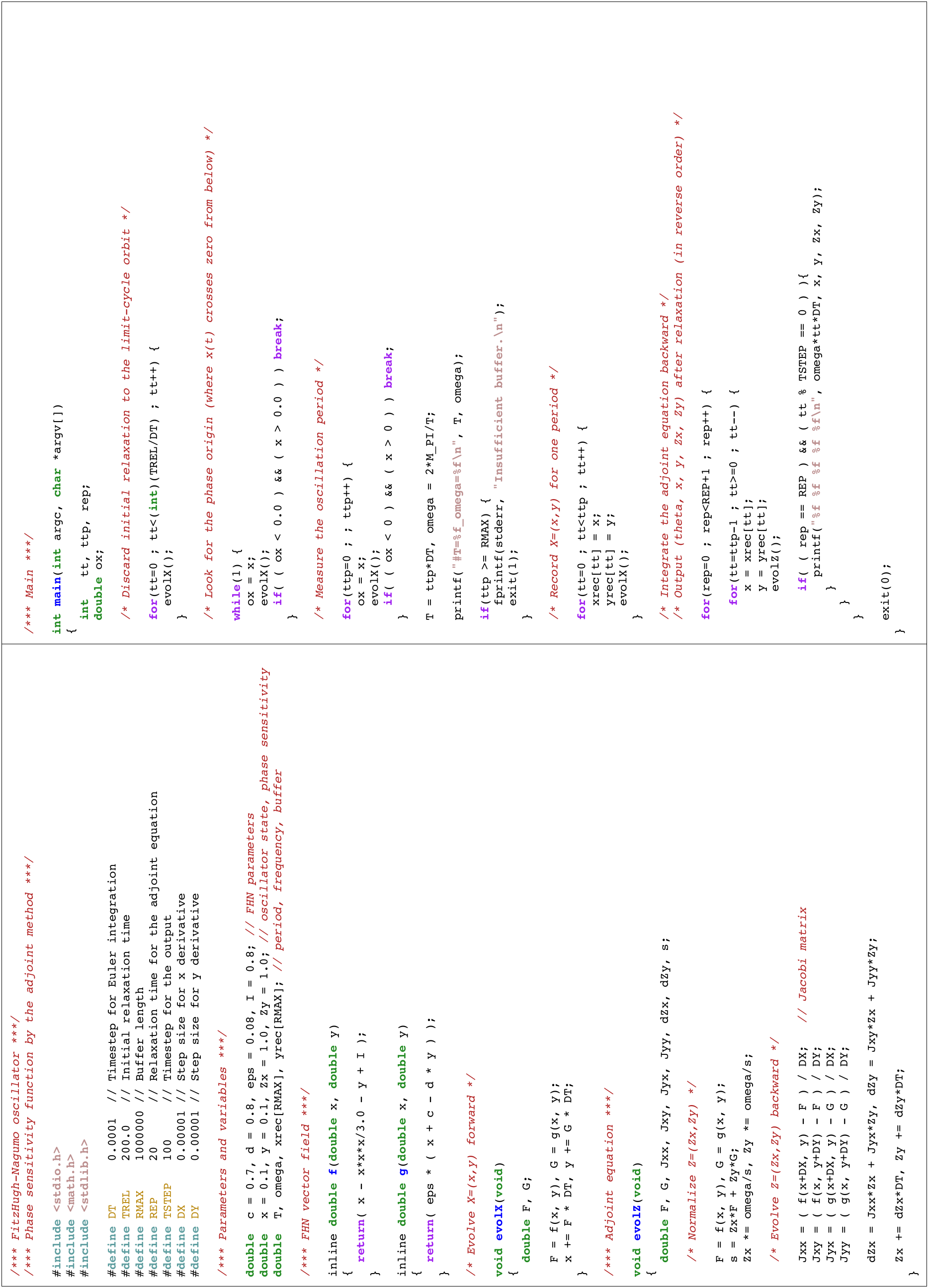}
	\caption{Sample source code in C for calculating the phase sensitivity function of the FitzHugh-Nagumo oscillator.}
	\label{fig-adjoint}
\end{figure}

Figure~\ref{fig-adjoint} shows an example source code in C that calculates the phase sensitivity function of the FitzHugh-Nagumo oscillator using the adjoint method. For readability of the code, numerical integration is performed by the simplest Euler integration.

\end{document}